\documentclass{emulateapj}
\usepackage{natbib}
\usepackage{graphicx}
\usepackage{epstopdf}
\usepackage[english]{babel}
\usepackage[colorlinks,linkcolor={blue},citecolor={blue},urlcolor={blue}]{hyperref}
\def\mpchi{\,h^{-1}{\rm {Mpc}}}

\def\msun{\,h^{-1}{\rm M_\sun}}

\def\be{\begin{equation}}
\def\ee{\end{equation}}
\def\ba{\begin{eqnarray}}
\def\ea{\end{eqnarray}}

\shorttitle{Luminosity and Color Dependent Clustering} \shortauthors{Guo et al.}

\begin{document}

\title{THE CLUSTERING OF GALAXIES IN THE SDSS-III BARYON OSCILLATION
SPECTROSCOPIC SURVEY: LUMINOSITY AND COLOR DEPENDENCE AND REDSHIFT EVOLUTION}

\author{
Hong Guo${}^{1,2}$, Idit Zehavi${}^{1}$, Zheng Zheng${}^{2}$, David H.
Weinberg${}^{3}$, Andreas A. Berlind${}^{4}$, Michael Blanton${}^{5}$, Yanmei
Chen${}^{6}$, Daniel J. Eisenstein${}^{7}$, Shirley Ho${}^{8,9}$, Eyal
Kazin${}^{10}$, Marc Manera${}^{11}$, Claudia Maraston${}^{11,12}$, Cameron
K. McBride${}^{7}$, Sebasti\'an E. Nuza${}^{13}$, Nikhil
Padmanabhan${}^{14}$, John K. Parejko${}^{14}$, Will J. Percival${}^{11}$,
Ashley J. Ross${}^{11}$, Nicholas P. Ross${}^{8}$, Lado Samushia${}^{11,15}$,
Ariel G. S\'anchez${}^{16}$, David J. Schlegel${}^{8}$, Donald P.
Schneider${}^{17,18}$, Ramin A. Skibba${}^{19}$, Molly E. C. Swanson${}^{7}$,
Jeremy L. Tinker${}^{5}$, Rita Tojeiro${}^{11}$, David A. Wake${}^{20,21}$,
Martin White${}^{8,22,23}$, Neta A. Bahcall${}^{24}$, Dmitry
Bizyaev${}^{25}$, Howard Brewington${}^{25}$, Kevin Bundy${}^{26}$, Luiz N.
A. da Costa${}^{27,28}$, Garrett Ebelke${}^{25}$, Elena
Malanushenko${}^{25}$, Viktor Malanushenko${}^{25}$, Daniel Oravetz${}^{25}$,
Graziano Rossi${}^{29,30}$, Audrey Simmons${}^{25}$, Stephanie
Snedden${}^{25}$, Alina Streblyanska${}^{31,32}$, and Daniel
Thomas${}^{11,12}$}

\affil{${}^{1}$ Department of Astronomy, Case Western Reserve University, OH
44106, USA}

\affil{${}^{2}$ Department of Physics and Astronomy, University of Utah, UT
84112, USA}

\affil{${}^{3}$ Department of Astronomy and CCAPP, Ohio State University,
Columbus, OH 43210, USA}

\affil{${}^{4}$ Department of Physics and Astronomy, Vanderbilt University,
Nashville, TN 37235, USA}

\affil{${}^{5}$ Center for Cosmology and Particle Physics, New York
University, New York, NY 10003, USA}

\affil{${}^{6}$ Department of Astronomy, Nanjing University, Nanjing 210093,
China}

\affil{${}^{7}$ Harvard-Smithsonian Center for Astrophysics, 60 Garden St.,
Cambridge, MA 02138, USA}

\affil{${}^{8}$ Lawrence Berkeley National Laboratory, 1 Cyclotron Road,
Berkeley, CA 94720, USA}

\affil{${}^{9}$ Department of Physics, Carnegie Mellon University, 5000
Forbes Avenue, Pittsburgh, PA 15213, USA}

\affil{${}^{10}$ Center for Astrophysics and Supercomputing, Swinburne
University of Technology, P.O. Box 218, Hawthorn, Victoria 3122, Australia}

\affil{${}^{11}$ Institute of Cosmology \& Gravitation, Dennis Sciama
Building, University of Portsmouth, Portsmouth, PO1 3FX, UK}

\affil{${}^{12}$ SEPnet, South East Physics Network, www.sepnet.ac.uk}

\affil{${}^{13}$ Leibniz-Institut f\"ur Astrophysik Potsdam (AIP), An der
Sternwarte 16, D-14482 Potsdam, Germany}

\affil{${}^{14}$ Department of Physics, Yale University, 260 Whitney Ave, New
Haven, CT 06520, USA}

\affil{${}^{15}$ National Abastumani Astrophysical Observatory, Ilia State
University, 2A Kazbegi Ave., GE-1060 Tbilisi, Georgia}

\affil{${}^{16}$ Max-Planck-Institut f\"ur extraterrestrische Physik,
Postfach 1312, Giessenbachstr., D-85748 Garching, Germany}

\affil{${}^{17}$ Department of Astronomy and Astrophysics, The Pennsylvania
State University, University Park, PA 16802, USA}

\affil{${}^{18}$  Institute for Gravitation and the Cosmos, The Pennsylvania
State University, University Park, PA 16802, USA}

\affil{${}^{19}$ Center for Astrophysics and Space Sciences, University of
California, 9500 Gilman Drive, San Diego, CA 92093, USA}

\affil{${}^{20}$ Department of Astronomy, Yale University, New Haven, CT
06520, USA}

\affil{${}^{21}$ Department of Astronomy, University of Wisconsin-Madison,
475 N. Charter Street, Madison, WI, 53706, USA}

\affil{${}^{22}$ Department of Physics, University of California, 366 LeConte
Hall, Berkeley, CA 94720, USA}

\affil{${}^{23}$ Department of Astronomy, 601 Campbell Hall, University of
California at Berkeley, Berkeley, CA 94720, USA}

\affil{${}^{24}$ Princeton University Observatory, Princeton, NJ 08544, USA}

\affil{${}^{25}$  Apache Point Observatory, P.O. Box 59, Sunspot, NM
88349-0059, USA}

\affil{${}^{26}$ Kavli Institute for the Physics and Mathematics of the
Universe, University of Tokyo, Kashiwa, 277-8582, Japan}

\affil{${}^{27}$ Observat\'orio Nacional, Rua Gal. Jos\'e Cristino 77, Rio de
Janeiro, RJ - 20921-400, Brazil}

\affil{${}^{28}$ Laborat\'orio Interinstitucional de e-Astronomia - LineA,
Rua Gal. Jos\'e Cristino 77, Rio de Janeiro, RJ - 20921-400, Brazil}

\affil{${}^{29}$  CEA, Centre de Saclay, Irfu/SPP, F-91191 Gif-sur-Yvette,
France}

\affil{${}^{30}$  Paris Center for Cosmological Physics (PCCP) and
Laboratoire APC, Universit\'e Paris 7, F-75205 Paris, France}

\affil{${}^{31}$ Instituto de Astrofisica de Canarias (IAC), E-38200 La
Laguna, Tenerife, Spain}

\affil{${}^{32}$ Dept. Astrofisica, Universidad de La Laguna (ULL), E-38206
La Laguna, Tenerife, Spain}

\begin{abstract}
We measure the luminosity and color dependence and the redshift evolution of
galaxy clustering in the Sloan Digital Sky Survey-III Baryon Oscillation
Spectroscopic Survey  Ninth Data Release.  We focus on the projected
two-point correlation function (2PCF) of subsets of its CMASS sample, which
includes about $260,000$ galaxies over ${\sim}3,300\deg^2$ in the redshift
range $0.43<z<0.7$. To minimize the selection effect on galaxy clustering, we
construct well-defined luminosity and color subsamples by carefully
accounting for the CMASS galaxy selection cuts. The 2PCF of the whole CMASS
sample, if approximated by a power-law, has a correlation length of $r_0 =
7.93\pm0.06h^{-1}{\rm Mpc}$ and an index of $\gamma=1.85\pm0.01$. Clear
dependences on galaxy luminosity and color are found for the projected 2PCF
in all redshift bins, with more luminous and redder galaxies generally
exhibiting stronger clustering and steeper 2PCF. The color dependence is also
clearly seen for galaxies within the red sequence, consistent with the
behavior of SDSS-II main sample galaxies at lower redshifts. At a given
luminosity (k+e corrected), no significant evolution of the projected 2PCFs
with redshift is detected for red sequence galaxies. We also construct galaxy
samples of fixed number density at different redshifts, using
redshift-dependent magnitude thresholds. The clustering of these galaxies in
the CMASS redshift range is found to be consistent with that predicted by
passive evolution. Our measurements of the luminosity and color dependence
and redshift evolution of galaxy clustering will allow for detailed modeling
of the relation between galaxies and dark matter halos and new constraints on
galaxy formation and evolution.
\end{abstract}


\section{Introduction}

Galaxies in the universe are observed to display a wide range of properties,
such as luminosity, color, stellar mass, age, morphology, and spectral type.
These properties encode information about galaxy formation and evolution and
are related to the environment hosting the galaxies. Different populations of
galaxies are thus expected to trace the underlying dark matter distribution
in different ways.

Contemporary galaxy redshift surveys, most notably the Sloan Digital Sky
Survey (SDSS; \citealt{York00}), have transformed the study of large-scale
structure, enabling pristine measurements and detailed studies of galaxy
clustering. Galaxy clustering provides a powerful approach to probe the
complex relation between galaxies and the underlying dark matter distribution
\citep[e.g.,][]{Kaiser84}. The dependence of galaxy clustering on galaxy
properties has been observed in numerous galaxy surveys
\citep[e.g.,][]{Davis76,Davis88,Hamilton88,Loveday95,Benoist96,Guzzo97,Norberg01,Norberg02,
Zehavi02,Zehavi05b,Zehavi11,Budavari03,Madgwick03,Li06,
Coil06,Coil08,Meneux06,Meneux08,Wang07,Wake08,Swanson08,Meneux09,Ross09,Skibba09b,
Loh10,Ross10,Ross11b,Wake11,Christodoulou12,Mostek12}.

In general, more luminous and redder galaxies are found to be more strongly
clustered than their fainter and bluer counterparts. Similarly, early-type
(elliptical) galaxies exhibit stronger clustering than late-type (spiral)
ones. Galaxy luminosity and color are perhaps the two major readily observed
properties, which also facilitate comparison between different surveys and
are less dependent on the stellar evolution models. Moreover, they have
proven to be the two properties most predictive of galaxy environment, such
that any residual dependence on morphology or surface brightness is weak
\citep{Blanton05}. In this paper, we measure the dependence of galaxy
clustering on color and luminosity for galaxies in the SDSS-III Baryon
Oscillation Spectroscopic Survey (BOSS; \citealt{Eisenstein11,Dawson13}) and
study the implications.

Galaxy clustering measurements, once properly interpreted, can provide key
information about galaxy formation and evolution. In particular, the
theoretical understanding of galaxy clustering has been greatly advanced with
the development of dark matter halo models (see \citealt{Cooray02} and
references therein). In the cosmological constant + cold dark matter
($\Lambda$CDM) paradigm, galaxies form and evolve in dark matter halos. While
the properties and clustering of dark matter halos are well understood with
the help of analytic models and numerical simulations
\cite[e.g.,][]{Mo96,Springel05}, the properties of galaxies are hard to
predict because of the complex baryonic processes and the lack of complete
theory of galaxy formation.

Galaxy clustering offers an opportunity to connect galaxies to dark matter
halos, providing a new direction in studying galaxy formation and evolution.
The halo occupation distribution (HOD) framework \citep[see
e.g.,][]{Peacock00,Seljak00,Scoccimarro01,Berlind02,Berlind03,Zheng05} or the
closely related conditional luminosity function (CLF) method
\citep{Yang03,Yang05} describe the number of galaxies as a function of halo
mass, and galaxy clustering is used to constrain the HOD or CLF parameters.
The subhalo abundance matching method makes use of subhalos in high
resolution $N$-body simulations and connects them to galaxies to interpret
galaxy clustering \citep[see, e.g.,][]{Kravtsov04,Conroy06, Guo10,Nuza12}.
Such models essentially convert galaxy clustering measurements to the
relation between galaxies and halos, which provides strong tests of galaxy
formation models.

The galaxy-halo connections inferred at different redshifts, together with
the theoretically known halo evolution, can lead to empirical constraints on
galaxy evolution. For example, \cite{Zheng07} compare the HODs for $z{\sim}1$
DEEP2 and $z{\sim}0$ SDSS galaxies and find a halo mass dependent growth of
stellar mass of central galaxies, separating into contributions from star
formation and mergers. \cite{Tinker10} analyze four samples of galaxies from
redshift $z=0.4$ to $z=2.0$. They find that more than $75\%$ of the red
satellite galaxies move onto the red sequence because of halo mergers, while
the mechanism for central galaxies to move to the red sequence evolves from
$z=0.5$ to $z=0$.

To advance our understanding of galaxy evolution, improved measurements of
galaxy clustering in large galaxy surveys at different redshifts are
necessary. The color and luminosity dependence of clustering has been studied
in detail at $z{\sim}1$ for DEEP2 galaxies \citep{Coil08} and at $z{\sim}0$
for SDSS galaxies (\citealt{Zehavi11}; denoted Z11 hereafter). The trends
with color and luminosity are generally similar. However, while \cite{Coil08}
find no changes of the clustering within the red sequence, Z11 find a
continuous trend (in both amplitude and slope) of stronger clustering with
color in SDSS galaxies. This discrepancy may be caused by different sample
selections, but it may also be a signature of galaxy evolution (e.g., related
to the buildup of red sequence). Studying a sample from an intermediate
redshift range can provide new important information and a better
understanding of galaxy evolution.

In this paper, we use the recently released CMASS sample of the BOSS Data
Release 9 (DR9; \citealt{Ahn12}) to measure the clustering of galaxies in the
redshift range $0.43<z<0.7$ and study the dependence on galaxy luminosity and
color. The sample is constructed to contain a roughly volume-limited set of
massive and luminous galaxies (with a typical stellar mass of
$10^{11.3}\msun$; \citealt{Maraston12}) in this redshift range. This sample
has recently been used to accurately measure the baryon acoustic oscillation
(BAO) signature (see \citealt{Weinberg12} for a recent comprehensive review)
on large scales (${\sim}100\mpchi$) by \citet{Anderson12}. The sample has
been thoroughly vetted, and the robustness of the results and cosmological
constraints from the BAO and redshift-space distortions are explored in a
series of papers \citep{Reid12,Ross12,Ross13,Sanchez12,Samushia13,
Scoccola12,Tojeiro12a,Tojeiro12b}. The smaller scale clustering measurements
and HOD fits are first presented by \citet{White11} (for an earlier smaller
sample) and \citet{Nuza12}. Here we study the small to intermediate scale
(0.05--25$\mpchi$) two-point correlation functions (2PCFs) of CMASS galaxies,
focusing especially on the dependence on luminosity and color and the
implications on galaxy evolution from simple models. We will study the
evolution of CMASS galaxies based on HOD modeling in a companion paper.

The structure of the paper is organized as follows. In
Section~\ref{sec:data}, we briefly describe the CMASS sample and our method
of measuring the 2PCFs. The division to specific subsamples, the clustering
measurements and detailed dependence on luminosity and color, and the
implications for galaxy evolution are presented in Section~\ref{sec:results}.
We summarize our results in Section~\ref{sec:conclusion}. Appendix A
discusses the effect of different stellar evolution models on our results,
and Appendix B explores the robustness of the jackknife error estimates used.

Throughout the paper, we assume a spatially flat $\Lambda$CDM cosmology as in
\citet{Anderson12}, with $\Omega_m=0.274$, $h=0.7$, $\Omega_bh^2=0.0224$,
$n_s=0.95$, and $\sigma_8=0.8$, consistent with the best-fit model from the
Wilkinson Microwave Anisotropy Probe 7-year data \citep{Komatsu11}.

\section{Observations and Methods}
\label{sec:data}

\subsection{Data}
\label{subsec:data}

As part of the SDSS-III survey, BOSS selects luminous galaxies from the
multiple-band SDSS imaging \citep{Fukugita96,Gunn98,Gunn06,York00} for
spectroscopic observation to probe the large-scale BAO signals
\citep{Anderson12}. \cite{Dawson13} provide a comprehensive overview of BOSS,
while the technical details of BOSS are presented in \cite{Smee12} and
\cite{Bolton12}. The selection of BOSS galaxies is a union of targets in two
different redshift intervals. One is an extension of the SDSS-I/II Luminous
Red Galaxy (LRG) sample \citep{Eisenstein01}, referred to as LOWZ with
$0.2<z<0.43$ \citep{Parejko13}. The other, denoted as CMASS
\citep{White11,Anderson12}, includes ${\sim}260,000$ galaxies in DR9, and is
approximately stellar-mass limited at higher redshifts ($0.43<z<0.7$) with an
effective volume of ${\sim} 2.2\rm{Gpc^3}$ and an effective area of about
$3,300\deg^2$. In this paper, our study focuses on the CMASS sample. The
target selection cuts for CMASS galaxies are fainter and bluer than the LRG
sample in order to achieve a higher number density of about $3\times10^{-4}
h^3{\rm Mpc}^{-3}$ and sample a wider range of galaxies. The detailed target
selection of the CMASS sample is described in \cite{Padmanabhan13}, and
summarized in \cite{Eisenstein11} and \cite{Anderson12}. We briefly describe
here the major selection cuts that will affect our analysis of luminosity and
color dependence of the 2PCF.

The CMASS sample aims at selecting galaxies following a roughly constant
stellar mass cut at redshift $z>0.4$. The selection criteria of the CMASS
galaxies are defined by,
\begin{eqnarray}
  17.5 < i_{\rm cmod} &<& 19.9\label{eq:fluxcut} \\
  d_\perp &>& 0.55,\label{eq:dpcut} \\
  i_{\rm cmod} &<& 19.86 + 1.6(d_{\perp} - 0.8)\label{eq:slidecut} \\
  i_{\rm fib2} &<& 21.5 \\
  r_{\rm mod} - i_{\rm mod} &<& 2.0
\end{eqnarray}
where $d_\perp$ is defined as
\begin{equation}
  d_\perp = r_{\rm mod} - i_{\rm mod} - (g_{\rm mod} - r_{\rm mod})/8
\end{equation}
and all magnitudes are extinction corrected \citep{Schlegel98} and are in the
observed frame. While the magnitudes are calculated using {\tt cmodel}
magnitudes (denoted by the subscript ``cmod''), the colors are computed using
{\tt model} magnitudes (denoted by the subscript ``mod''; see
\citealt{Anderson12} for details). The magnitude $i_{\rm fib2}$ corresponds
to the $i$-band flux within the $2''$ fiber size. CMASS objects also pass
specific star-galaxy separation cuts, as described in \cite{Anderson12}.

\subsection{Selection Cuts and Sample Completeness}
\label{subsec:incomp}

\begin{figure*}
\epsscale{1.0}\plotone{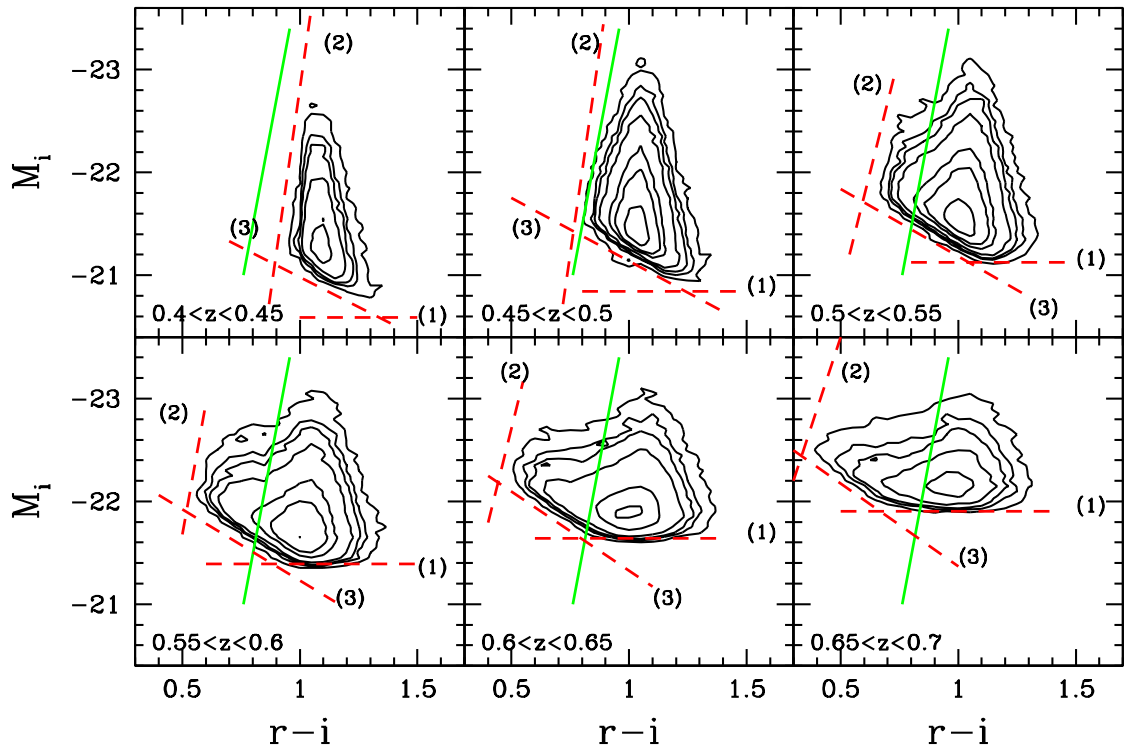} \caption{Color-magnitude diagram (CMD) of
CMASS galaxies in different redshift intervals. Both magnitude and color are
k+e corrected to $z=0.55$. The contours represent the number density
distribution of CMASS galaxies in the CMD. The approximate positions of the
selection cuts are shown as dashed lines in each panel, labeled with the
corresponding equation numbers of the selection cuts (see the text). The
green solid lines are our color cut (Equation~\ref{eq:colorcut}) for red and
blue galaxies. (A color version of this figure is available in the online
journal.)} \label{fig:cmd}
\end{figure*}
The CMASS galaxies are chosen by applying complex target selection cuts, and
it is thus difficult to construct exact volume-limited samples. Since we
intend to investigate the luminosity and color dependence of galaxy
clustering in this paper, we must pay particular attention to the
completeness in luminosity and color. With this goal in mind, we first
investigate the target selection cuts projected to the color-luminosity plane
at different redshifts, which will provide the appropriate boundaries in
constructing our samples.

Figure~\ref{fig:cmd} shows the color-magnitude diagram (CMD) at six narrow
redshift ranges with $\Delta z=0.05$. The contours represent the number
density distribution of CMASS galaxies in the CMD. The absolute magnitude
$M_i$ and $r-i$ color are k+e corrected to $z=0.55$ throughout the paper
using a global Flexible Stellar Population Synthesis (FSPS) model
\citep{Conroy09,Conroy10,Tojeiro11,Tojeiro12b}. Using other stellar evolution
models slightly changes the resulting magnitudes and colors, but does not
affect our analysis of galaxy clustering, as we discuss in
Appendix~\ref{app:models}. In each panel of Figure~\ref{fig:cmd}, only the
approximate positions of the selection cuts are shown as dashed lines, since
the selection cuts are made in apparent magnitudes and in three bands ($g,
r,$ and, $i$) while our CMD is shown for $r-i$ color and $M_i$ magnitude. The
slopes of the dashed lines follow the boundaries seen in the CMD contours in
each redshift bin. We focus on the three main cuts
(Equations~\ref{eq:fluxcut}--\ref{eq:slidecut}). Following \cite{Zehavi11},
we adopt a luminosity-dependent color cut to separate red and blue galaxies
(discussed in detail in Section~\ref{subsec:color}),
\begin{equation}
(r-i)_{\rm cut}=0.679-0.082(M_i+20), \label{eq:colorcut}
\end{equation}
represented by the green lines in Figure~\ref{fig:cmd}.

The horizontal cut in each panel corresponds to the $i$-band faint-end flux
limit (Equation~\ref{eq:fluxcut}, $i<19.9$), which selects galaxies brighter
than the corresponding absolute magnitude at each redshift. The slightly
tilted vertical dashed line on the left of the distribution is the $d_\perp$
cut (Equation~\ref{eq:dpcut}), which removes galaxies with bluer colors and
with lower redshifts \citep{Cannon06,Padmanabhan13}. The bottom-left dashed
line is the $i$-band sliding cut (Equation \ref{eq:slidecut}), which excludes
the fainter and bluer (thus lower stellar mass) galaxies from the sample.
These three main cuts evolve with redshift. A consequence of the $d_\perp$
and sliding cuts is that there are more blue galaxies at higher redshifts. At
$z>0.55$, while the CMASS sample is less affected by the $d_\perp$ cut, the
sliding cut must be carefully taken into account at all redshifts in
constructing complete galaxy samples. At $z<0.55$, blue galaxies are highly
incomplete (see an estimation of the fraction of star-forming galaxies in
CMASS in Figure~16 of \citealt{Chen12}). At $z<0.45$, the $d_\perp$ cut leads
to incompleteness even for the most luminous red galaxies.

A more sophisticated method to study the sample completeness would be to
simulate the galaxy properties (stellar mass, luminosity and color) at
different redshifts by assuming certain galaxy stellar evolution models and
to apply the selection cuts to simulated galaxies \citep{Swanson13}. While it
would generally depend on the assumptions in the galaxy evolution models,
such a study can provide a quantitative estimate of the sample completeness
as a function of color and luminosity. In this paper, however, we do not
intend to evoke such a model in our clustering analysis. Instead, we
empirically use the CMD and selection cuts at each redshift to construct
approximately complete galaxy samples. As we proceed in our analysis, we keep
in mind the boundaries of ``completeness'' defined by the selection cuts when
constructing our galaxy samples. One advantage of such an empirical method is
that it is largely model-independent.

\subsection{Clustering Measurements}
\label{sec:method}

In this paper, we focus our discussion on the galaxy 2PCF. We use the
Landy-Szalay estimator \citep{Landy93} to measure the 2PCF of galaxies,
\begin{equation}
\xi(r)=\frac{\rm{DD}-2\rm{DR}+\rm{RR}}{\rm{RR}}\label{eq:estimator}
\end{equation}
where $\rm{DD}$, $\rm{DR}$ and $\rm{RR}$ are the data--data, data--random,
and random--random pair counts measured from the data of $N$ galaxies and
random samples consisting of $N_R$ random points. These pair counts are
appropriately normalized by $N(N-1)/2$, $NN_R$, and $N_R(N_R-1)/2$,
respectively.

We measure the three-dimensional (3D) 2PCF $\xi(r_p,r_\pi)$, where $r_p$ and
$r_\pi$ are the separations of galaxy pairs perpendicular and parallel to the
line of sight. The redshift-space 2PCF $\xi(r_p,r_\pi)$ differs from the
real-space one because of redshift distortions induced by galaxy peculiar
velocities. The redshift distortions can be mitigated by projecting the 2PCF
along the line-of-sight direction, with the projected 2PCF $w_p(r_p)$
\citep{Davis83} defined and measured as
\begin{eqnarray}
w_p(r_p)=2\int_0^\infty \xi(r_p,r_\pi)dr_\pi = 2\sum_i\xi(r_{p},r_{\pi,i})\Delta r_{\pi,i}\label{eq:wp}
\end{eqnarray}
where $r_{\pi,i}$ and $\Delta r_{\pi,i}$ are the $i$--th bin of the
line-of-sight separation and its corresponding bin size. In practice, we sum
$\xi(r_p,r_\pi)$ along the line-of-sight direction up to $r_{\pi, {\rm
max}}=80\mpchi$ to include most of the correlated pairs. As our analysis
focuses on 2PCF measurements up to $r_p\sim 25\mpchi$, the clustering
measurements do not depend significantly on the assumed $r_{\pi, {\rm max}}$
once it is sufficiently larger. For example, if we integrate the
line-of-sight direction to $200\mpchi$, the contribution from
$80\mpchi<r_\pi<200\mpchi$ to $w_p(r_p)$ only introduces noisy fluctuations
of about $2\%$.

The projected 2PCF can be related to the real-space correlation function,
$\xi(r)$, by
\begin{equation}
w_p(r_p)
= 2 \int_{r_p}^{\infty} r\, dr\, \xi(r) (r^2-r_p^2)^{-1/2}
\label{eq:wp2}
\end{equation}
\citep{Davis83}. It is common practice to characterize $\xi(r)$ by a power
law on small scales, $\xi(r)=(r/r_0)^{-\gamma}$, and in such a case
$w_p(r_p)$ can be expressed as
\begin{eqnarray}
w_p(r_p)=r_p\left(r_0\over r_p\right)^\gamma\ \Gamma\left({1\over2}\right)\Gamma\left({\gamma-1\over2}\right)\left/
\Gamma\left({\gamma\over2}\right)\right . \label{eq:wppow}
\end{eqnarray}

The covariance matrix of the correlation function is estimated using the
jackknife resampling method (following \citealt{Zehavi05b,Zehavi11})
\begin{equation}
{\rm Cov}(\xi_i,\xi_j)={N-1 \over N}\sum_{l=1}^N (\xi_i^l-\bar{\xi_i})(\xi_j^l-\bar{\xi_j})
\label{eq:jk}
\end{equation}
where $N\equiv100$ is the total number of jackknife samples, $\xi_i^l$ is the
2PCF in the $i$--th pair separation bin measured from the $l$--th jackknife
sample, and $\bar{\xi_i}$ is the average over all samples. We expand on the
tests performed by \cite{Zehavi05b} and investigate the accuracy of the
jackknife error estimates using CMASS mock catalogs and find good agreement
between the jackknife estimates and those from multiple mocks (see
Appendix~\ref{app:jack} for more details).

We focus on analyzing our measurement of the galaxy 2PCF on the small to
intermediate scales ($0.05\mpchi<r_p<25\mpchi$), aiming at probing the
relation between galaxies and their host halos. However, the small-scale
measurement of the 2PCF is limited by the fiber collision effect in the
spectrograph system of SDSS-III, where two fibers on the same plate cannot be
placed closer than an angular separation of $62''$
\citep{Dawson13,Anderson12}. As a result, about $5.5\%$ of CMASS galaxies do
not have redshifts, strongly affecting small-scale clustering measurements.
We correct for this effect using the method proposed and tested by
\cite{Guo12}, which divides the galaxy sample into two distinct populations,
one free of fiber collisions (referred to as $\rm{D_1}$) and the other
consisting of potentially collided galaxies (denoted as $\rm{D_2}$). The
total clustering signal is a combination of the contributions from these two
populations, where the contribution of the collided population is estimated
from the resolved galaxies in tile-overlap regions.

As discussed in detail in \cite{Guo12}, there are two main systematics that
could impact the accuracy of such a method to treat fiber collisions. One is
possible density variations between the tile overlap and non-overlap regions,
which is found to be insignificant from tests with mock galaxy catalogs.
Another effect is that galaxies in collided triplets (or even higher-order
colliding groups) can only be fully recovered in regions covered by at least
three tiles, making the estimation from close pairs in two-tile regions not
accurate (because of the lack of $\rm{D_2D_2}$ close pairs). This effect is
alleviated by an additional correction term using the measured close pairs in
the two-tile regions. After full application of the \cite{Guo12} method, we
estimate the remaining systematic errors to be less than 3--5\%. It is
generally difficult to have an unbiased correction on all scales, but this
approach provides the best estimate of the galaxy 2PCF on small scales,
compared with other possible methods.

When counting pairs for the 2PCF, each galaxy is assigned a series of weights
to reduce variance in the measurements and take into account different
effects. Following \cite{Anderson12}, we apply a scale-independent weight to
optimize the clustering measurements \citep{Feldman94}
\begin{equation}
w_{\rm{FKP}}={1\over1+\bar{n}(z)P_0},
\end{equation}
where $\bar{n}(z)$ is the mean density at redshift $z$, and $P_0=2\times10^4
h^{-3}{\rm Mpc}^3$. This equation provides similar results to the minimum
variance weight used by \cite{Hamilton93} and \cite{Zehavi02}. Another
employed weight, $w_{\rm{rf}}$, accounts for the fact that not every galaxy
with a spectrum taken has a reliable redshift measurement. The ``redshift
failures'' are dependent on the positions of fibers on the plates
\citep{Ross12} and are corrected by up-weighting the nearest galaxy that has
an accurate redshift. The final weight, $w_{\rm{sys}}$, is caused by the
scarcity of galaxies detected due to foreground bright stars \citep{Ross11b}.
\cite{Ross12} present a comprehensive study of potential systematic effects
in the 2PCF analysis, and compute a set of weights $w_{\rm{sys}}$ based on
the stellar density and $i_{\rm{fib2}}$ magnitude. Therefore, the total
weight applied to each galaxy is
\begin{equation}
w_{\rm{tot}}=w_{\rm{FKP}}w_{\rm{sys}}w_{\rm{rf}}.
\end{equation}
The quantity $w_{\rm{tot}}$ is applied to both the $\rm{D_1}$ and $\rm{D_2}$
populations in the fiber collision correction. The systematic weight
$w_{\rm{sys}}$ only has a small effect on small and intermediate scales, but
significantly changes the clustering measurements on BAO scales.

We construct the random catalogs according to the detailed angular selection
of the DR9 galaxy sample. The radial selection function for each sample is
taken into account by assigning the shuffled galaxy redshifts to the random
objects. The shuffling method provides a better representation of the true
distribution compared with a smooth spline fit to the observed galaxy
redshift distribution, as detailed in \cite{Ross12}. This process is done
separately for the northern and southern Galactic Caps to account for the
different number density distributions \citep[see][for details]{Anderson12}.
To apply the fiber collision correction, separate random catalogs for the
$\rm{D_1}$ and $\rm{D_2}$ populations are used. Denoting the fraction of
recovered $\rm{D_2}$ galaxies as $N_2'/N_2$, for the $\rm{D_2}$ random
catalogs we apply an additional angular mask $N_2'/N_2$ in each sector (see
\citealt{Guo12} for details).

\section{Results}
\label{sec:results}

\subsection{2PCF of the Full CMASS Sample}\label{subsec:2pcf}
\begin{figure*}
\epsscale{1.0}\plotone{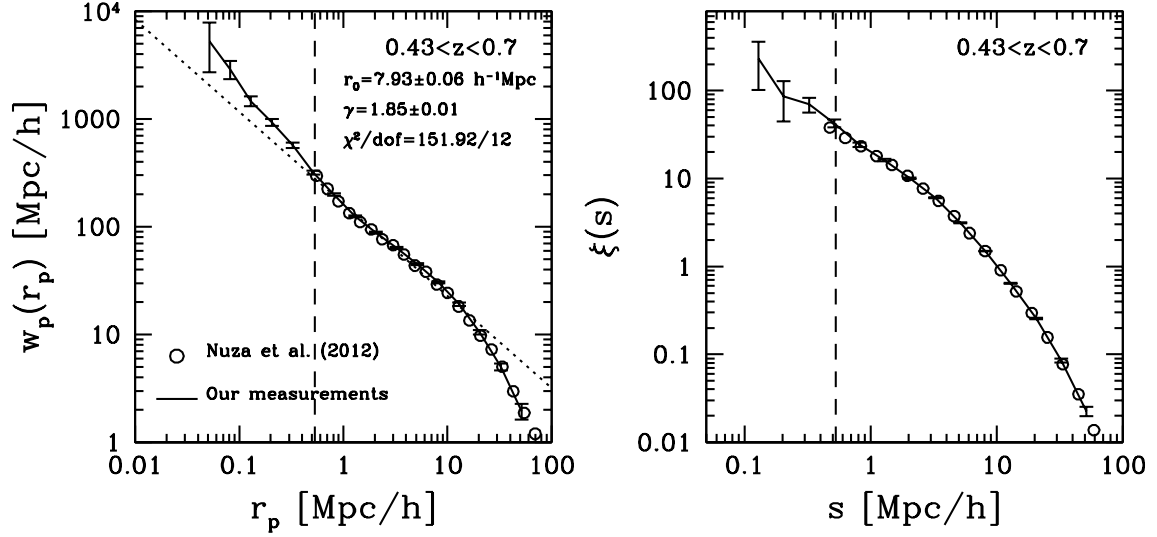}\caption{Projected (left panel) and
redshift-space(right panel) 2PCFs for the entire CMASS sample in the range of
$0.43<z<0.7$. The solid lines present our measurements. The open symbols are
the recent measurements of \cite{Nuza12}, which are in good agreement. The
vertical dashed lines correspond to the maximal fiber collision scale
(${\sim}0.53\mpchi$) for $z=0.7$. The dotted line in the left panel is a
power-law fit to $w_p$ for $0.05\mpchi<r_p<25\mpchi$ with the corresponding
parameters as labeled.} \label{fig:wpcmass}
\end{figure*}
Before presenting the luminosity and color dependence of the 2PCFs for CMASS
galaxies, we show in Figure~\ref{fig:wpcmass} the projected and
redshift-space 2PCFs for the entire CMASS sample in the redshift range
$0.43<z<0.7$ (solid lines). For comparison, we also display the recent CMASS
DR9 measurements of \citet[open symbols]{Nuza12}, which are limited to
slightly larger scales ($\gtrsim 0.5\mpchi$) and show good agreement on all
measured scales. The vertical dashed lines indicate the maximal fiber
collision scale, ${\sim}0.53\mpchi$ at redshift $z=0.7$. By applying the
fiber-collision correction method of \cite{Guo12}, we are able to robustly
measure the small-scale clustering (note the small error bars). This should
enable better constraints on the spatial distribution of galaxies inside dark
matter halos and a better constraint on the fraction of satellite galaxies,
which we will address in future work.

The dotted line in the left panel of Figure~\ref{fig:wpcmass} is the
power-law fit to $w_p$ for $0.05\mpchi<r_p<25\mpchi$, based on
Equation~\ref{eq:wppow}, using the full error covariance matrix. Although
$w_p$ clearly deviates from a power-law (also shown from the
$\chi^2/\rm{dof}$ of the fitting), a power-law fit provides a simple
characterization of the clustering and allows for easy comparisons among the
2PCFs of different galaxy samples. The correlation length for the CMASS
sample is $r_0=7.93\pm0.06\mpchi$ and the slope $\gamma=1.85\pm0.01$ (note
that given the large bestfit $\chi^2$, the error bars here should be
interpreted with care). These values are similar to clustering measurements
of LRGs at $z\sim0.55$ from the 2dF-SDSS LRG and QSO (2SLAQ) survey
\citep{Cannon06,Ross07,Wake08}, as expected, since the CMASS galaxy color
selections were based on the 2SLAQ LRG selection.

In order to study the large-scale galaxy bias, we focus on scales larger than
typical dark matter halo sizes. By fitting the ratio between the measured
galaxy $w_p$ and the theoretically computed dark matter $w_p$ on scales
$3\mpchi<r_p<25\mpchi$ (detailed in Section~\ref{subsec:bias}), we obtain a
``large-scale'' galaxy bias factor of $b=2.16\pm0.01$, consistent with the
measurements of other authors \citep{White11,Nuza12,Sanchez12,Shen12}. The
galaxy bias determined from the projected 2PCF has less scale dependence than
that from the redshift-space 2PCF. The exact value of the implied bias can
depend on the fitting scales and fitting methods. We find that if the minimal
fitting scale is changed from $3\mpchi$ to $5\mpchi$ (or larger), the
resulting bias only varies slightly at the $2\sigma$ level. In the right
panel of Figure~\ref{fig:wpcmass} we present the redshift-space 2PCF $\xi(s)$
on scales above $s>0.1\mpchi$, where it can be reliably measured. The fiber
collisions in that case impact larger scales than indicated by the dashed
line, since $s$ includes the contribution from the line-of-sight separations,
$s^2=r_p^2+r_\pi^2$ (see more discussion in \citealt{Guo12}).

\subsection{Luminosity Dependence}
\label{subsec:lum}
\subsubsection{Luminosity Cuts}
\label{subsubsec:lumcut}
\begin{figure*}
\epsscale{1.0}\plotone{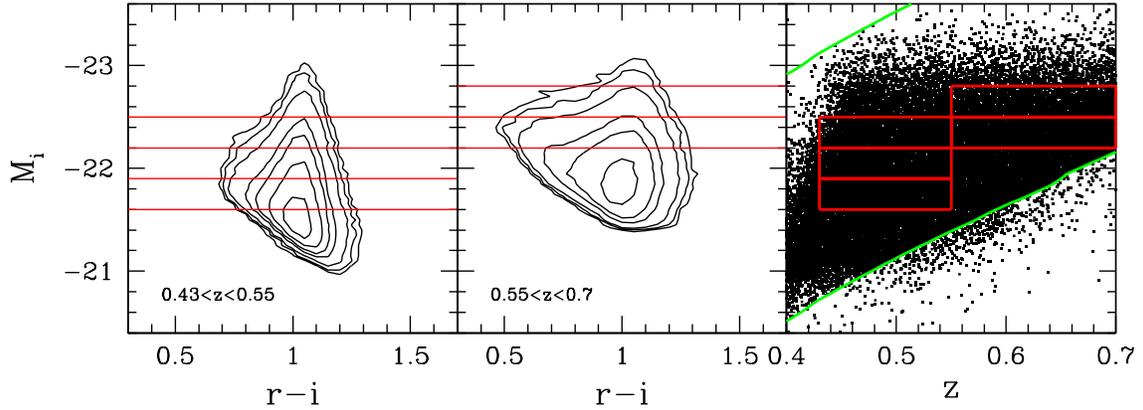}\caption{Color-magnitude diagrams of CMASS
galaxies in the two redshift bins we use, as well as overall distribution of
galaxies in $i$-band absolute magnitude and redshift. The red lines delineate
the luminosity bin samples we study. The two green lines in the right panel
represent the $i$-band flux limits of Equation~\ref{eq:fluxcut}, which is
also k+e corrected to $z=0.55$. (A color version of this figure is available
in the online journal.)} \label{fig:lumcuts}
\end{figure*}
We now investigate the luminosity dependence of CMASS galaxy clustering. To
minimize the influence of sample incompleteness, we carefully construct
samples of different luminosities by accounting for the selection cuts as a
function of redshifts discussed in Section \ref{subsec:incomp}. We divide
galaxies into two redshift bins, $0.43<z<0.55$ and $0.55<z<0.7$. The
color-magnitude distributions in these two redshift bins, the overall
magnitude-redshift distribution, and the cuts used to define our samples are
shown in Figure~\ref{fig:lumcuts}. In the right-most panel, some galaxies lie
below the faint flux limit (denoted by the lower green line), reflecting the
change between the photometry at the time of targeting and that from the
final processing. To keep a uniform criterion, we construct our luminosity
samples based on the targeting photometry \citep[see details
in][]{Anderson12}. We avoid the impact of the sliding cut by only selecting
galaxies brighter than the intersections between the $d_\perp$ and sliding
cuts. The incompleteness caused by the $d_\perp$ cut at $z<0.55$ cannot be
avoided. Such a limitation means that the blue galaxies are incomplete,
especially for the low-redshift samples, while the red galaxies are close to
complete for $z>0.5$, which is a caveat to remember when interpreting the
results. We will thus also study the luminosity dependence limited to the
more complete red galaxies. The sliding cut also impacts our ability to study
fainter galaxies at high redshift, resulting in the unsampled ``triangle''
region above $z=0.55$ in the right panel of Figure~\ref{fig:lumcuts}. We have
three and two luminosity bin samples at lower and higher redshifts,
respectively, each with a bin width of 0.3 magnitude, as shown in the figure.

\begin{deluxetable*}{llrrrcrrr}
\tablewidth{0pt} \tablecaption{\label{tab:lumsub}Samples of different
luminosities} \tablehead{\colhead{$M_i$ range} & \colhead{$z$ range} &
\colhead{$N_{\rm{tot}}$} & \colhead{$N_{\rm{red}}$} &
\colhead{$N_{\rm{blue}}$} & \colhead{$V_z [{\rm Gpc/h}]^3$} & \colhead{$r_0$}
& \colhead{$\gamma$} & \colhead{$\chi^2/\rm{dof}$}} \tablecomments{$r_0$ and
$\gamma$ are obtained from fitting a power-law to $w_p(r_p)$ using the full
error covariance matrices for $0.1\mpchi<r_p<25\mpchi$. The ratios between
$\chi^2$ and degrees-of-freedom (dof) of the fits are also shown.} \startdata
-21.9,\ -21.6 & 0.43,\ 0.55 & 48391 & 46235 & 2156 & 0.484 & $7.64\pm0.12$ & $1.86\pm0.02$ & 47.89/10\\
-22.2,\ -21.9 & 0.43,\ 0.55 & 24190 & 22419 & 1771 & 0.484 & $8.49\pm0.18$ & $1.91\pm0.03$ & 45.87/10\\
-22.5,\ -22.2 & 0.43,\ 0.55 & 9308  & 8687  & 621  & 0.484 & $9.99\pm0.31$ & $1.89\pm0.04$ & 10.22/10\\
-22.5,\ -22.2 & 0.55,\ 0.70 & 23404 & 15821 & 7583 & 0.851 & $8.56\pm0.19$ & $1.91\pm0.03$ & 10.68/10\\
-22.8,\ -22.5 & 0.55,\ 0.70 & 7484  & 5135  & 2349 & 0.851 & $10.40\pm0.32$ &
$1.88\pm0.06$ & 15.13/10
\enddata
\end{deluxetable*}
The total numbers of galaxies, $N_{\rm{tot}}$, and the comoving volume,
$V_z$, in each luminosity bin are shown in Table~\ref{tab:lumsub}. We also
provide the number of blue and red galaxies using the color cut
(Equation~\ref{eq:colorcut}). Power-law fits of Equation~\ref{eq:wppow} to
the projected 2PCF $w_p$ for $0.1\mpchi<r_p<25\mpchi$ are also given in the
table. It is evident that at lower redshifts, red galaxies dominate the
samples, and blue galaxies contribute less than $10\%$. At higher redshifts,
approximately $32\%$ of the sample are blue galaxies. This could still be
partly related to the CMASS selection effects. If the change in blue galaxy
fraction is caused by evolution, i.e., blue galaxies turning red with time,
we would expect the luminosity dependence of the 2PCF for red galaxies to
also evolve with redshift.

\subsubsection{The Dependence of Galaxy 2PCF on Luminosity}
\label{subsubsec:wplum}

\begin{figure*}
\epsscale{1.0}\plotone{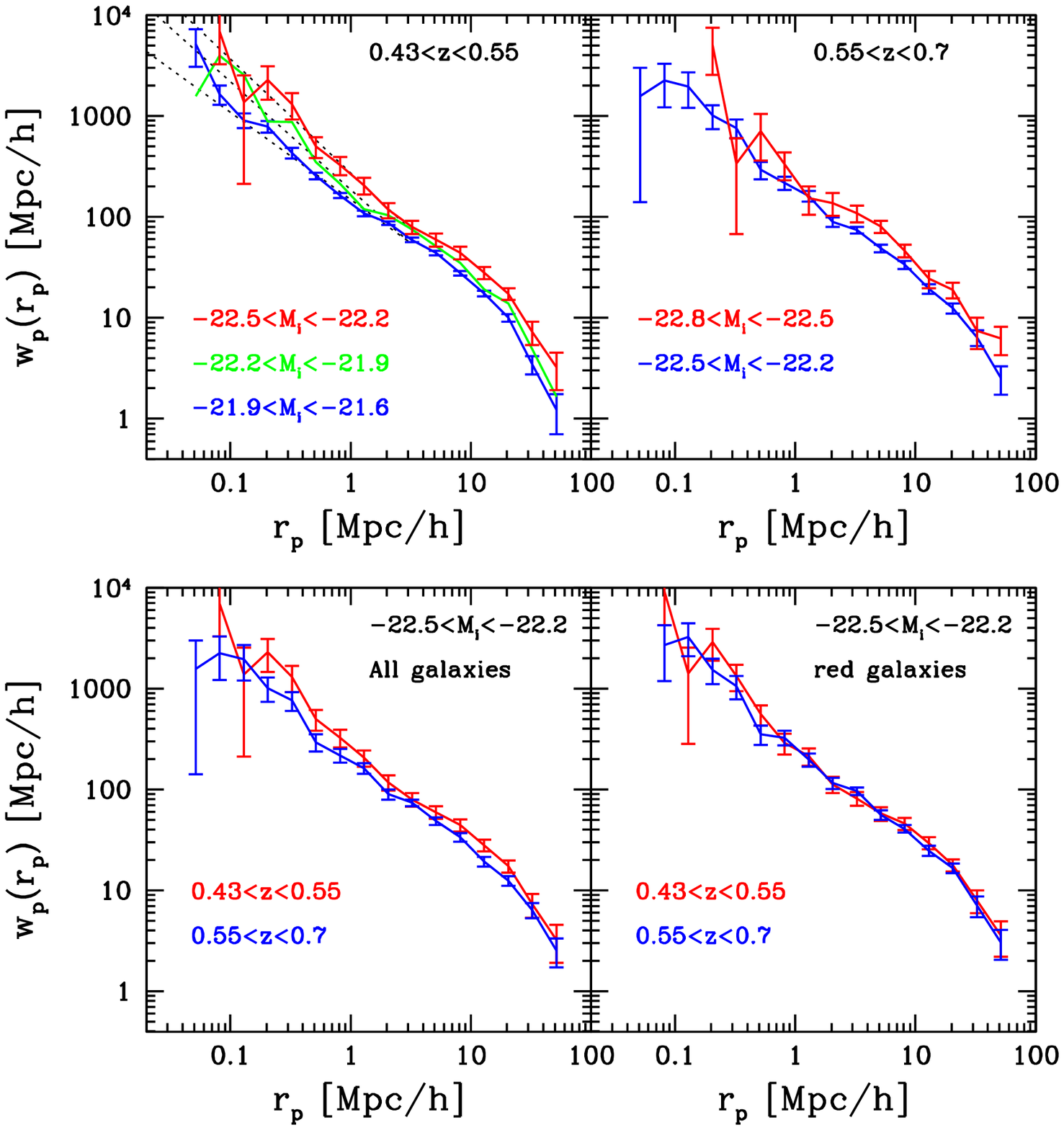}\caption{Projected correlation functions,
$w_p(r_p)$, for the various luminosity subsamples at low (top left) and high
redshift (top right). The bottom panels present the redshift evolution of
$w_p(r_p)$ in the luminosity interval $-22.8<M_i<-22.2$ for all the galaxies
in the sample (left) and only for the red galaxies (right). Error bars shown
are from the diagonal elements of the jackknife covariance matrices. The
dotted lines in the top left panel are the power-law fits to the $w_p$ in the
range of $0.1\mpchi<r_p<2\mpchi$ to provide a guide of the small-scale slope.
(A color version of this figure is available in the online journal.)}
\label{fig:wplum}
\end{figure*}
The projected 2PCFs of different luminosity samples at the two redshift bins
are shown in the top panels of Figure~\ref{fig:wplum}. At both redshifts, the
luminosity dependence of $w_p(r_p)$ is evident, with more luminous galaxies
exhibiting stronger clustering, consistent with the results of the SDSS-I/II
main sample \citep{Zehavi05b,Zehavi11,Li06}. At $z<0.55$, since red galaxies
contribute $90\%$ of the CMASS galaxy population, the luminosity dependence
mostly reflects the clustering environment of the red galaxies.  At $z>0.55$,
our measurements of the 2PCF become noisier because of the lower number of
galaxies. Even after accounting for the uncertainties in the measurements,
the luminosity dependence of clustering is significant in the redshift range
of CMASS galaxies.

According to the HOD modeling results in SDSS-I/II
\citep{Zehavi05b,Zehavi11}, the increase in the clustering amplitude for more
luminous samples reflects the shift in the host halo mass toward the high
mass end. The modeling results in \citet{Zehavi05b,Zehavi11} also show that
the satellite fraction drops as the luminosity of galaxies increases. Our
measurements naively seem to support such a result --- although the 2PCFs
becomes generally noisier for higher luminosity samples, the uncertainties in
the 2PCFs on small scales increase faster. At such scales, satellite galaxies
have a significant contribution to the clustering signal. Therefore, the
increase in the measurement errors could be a reflection of the lack of
satellites in more luminous samples.

The shapes of $w_p$ for all luminosity samples are similar. The deviation
from a power-law in $w_p$ is somewhat more apparent for brighter galaxies,
consistent with the results from main galaxies
\citep{Zehavi04,Zehavi05b,Zehavi11}. In the halo model, the slope of $w_p$
has a rapid change around a few Mpc, indicating the transition from
intra-halo galaxy pairs (1-halo term) to inter-halo galaxy pairs (2-halo
term). In our measurements, we see that this transition scale increases with
increasing luminosity, in agreement with the interpretation that more
luminous galaxies reside in more massive (hence larger) halos. The dotted
lines in the top left panel of Figure~\ref{fig:wplum} show the power-law fits
to $w_p$ in the range of $0.1\mpchi<r_p<2\mpchi$. There is an apparent weak
trend that brighter galaxies have a steeper slope in $w_p$ on small scales,
in line with the result in Z11. The dotted lines in the top left panel show
the power-law fits to the $w_p$ in the range of $0.1\mpchi<r_p<2\mpchi$, the
slope varies from $1.86\pm0.04$ (the faintest sample) to $2.13\pm0.11$ (the
brightest sample). The trend can also be interpreted as a result of the
change in the host halo mass scale (see Figure 7 and Appendix~A in
\citealt{Zheng09}).

The bottom panels of Figure~\ref{fig:wplum} show the redshift evolution of
the 2PCFs of galaxies in a fixed luminosity bin $-22.5<M_i<-22.2$. The left
panel is for all the galaxies (both blue and red). Galaxies at a lower
redshift appear to have a higher clustering strength. This result may be due
to the incompleteness of blue galaxies at lower redshifts, and the inclusion
of more (less clustered) blue galaxies at higher redshifts. We therefore also
examine the redshift evolution of the 2PCF in the $-22.5<M_i<-22.2$ sample
for red galaxies only, as shown in the bottom-right panel of
Figure~\ref{fig:wplum}, where the red galaxies are defined by the color cut
in Equation~\ref{eq:colorcut}. We find only slight evolution with redshift
for the red galaxies (at most $18\%$ in $w_p$ for $r_p>1\mpchi$), not
significant within the measurement errors, implying that the differences in
the sample of all galaxies (bottom-left panel) are mostly induced by the blue
galaxies.

In the redshift range of $0.16<z<0.44$, \cite{Zehavi05a} find no strong
evolution trend in $w_p(r_p)$ in the SDSS LRG sample. Their sample of
$-23.2<M_g<-21.8$ at $0.16<z<0.44$ has a comoving number density of
$n{\sim}0.2\times10^{-4}h^{3}{\rm Mpc}^{-3}$ (their Figure 2), similar to the
number density of our sample of $-22.5<M_i<-22.2$ at $0.43<z<0.7$ shown in
the bottom right panel of Figure~\ref{fig:wplum}. Combining their results
with ours, we would infer that there is no strong redshift evolution in $w_p$
of luminous red galaxies in the redshift range of $0.1<z<0.7$, consistent
with the results of \cite{Wake08}, implying that the effect of structure
growth roughly cancels that of evolution of galaxy bias. As will be discussed
in Section~\ref{subsec:bias}, within the errorbars, the trend is also roughly
consistent with passive evolution.

\subsection{Color Dependence} \label{subsec:color}
\subsubsection{Color Cuts}\label{subsubsec:colorcut}

\begin{figure*}
\epsscale{1.0}\plotone{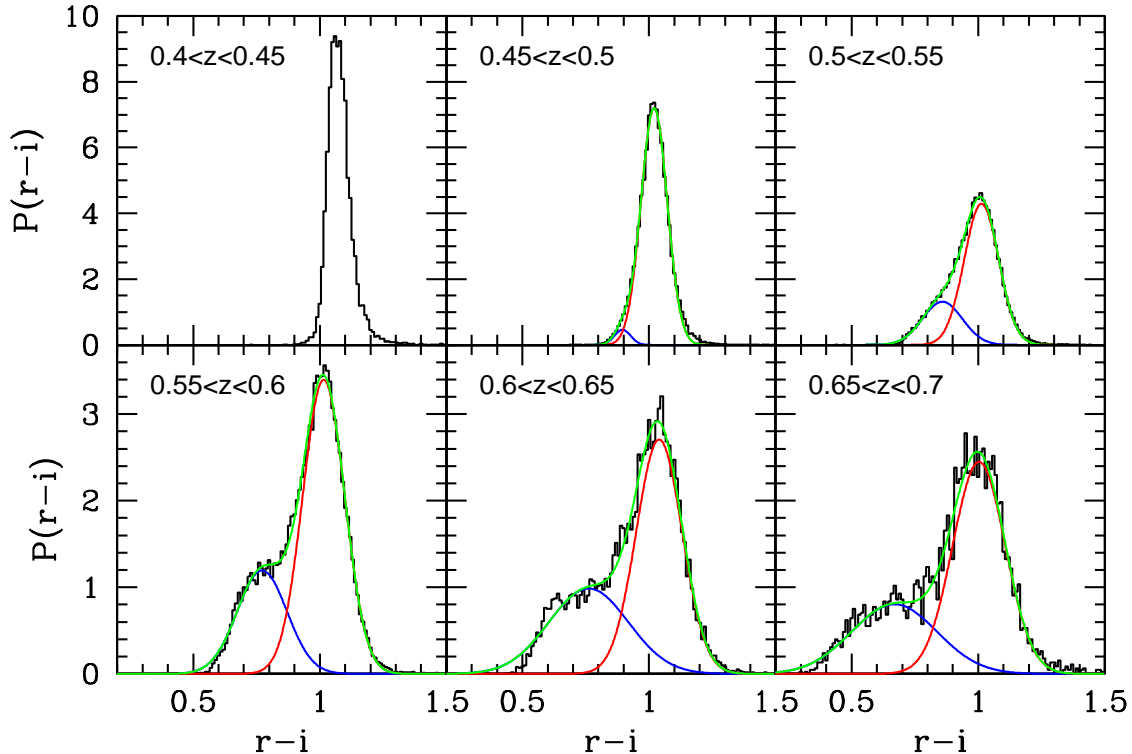}\caption{Probability distribution
function of $r-i$ color at different redshift slices, for CMASS samples of
high completeness (see text). The black lines are the histogram of $r-i$ for
all galaxies. The red and blue lines are the bimodality fitting using two
Gaussian distributions, with the green curves as their combination. We do not
fit the distribution of $r-i$ for $0.4<z<0.45$ because both the red and blue
galaxies are far from complete in this redshift interval. (A color version of
this figure is available in the online journal.) }\label{fig:histriz}
\end{figure*}
The various target selection cuts make it difficult to discern the ``red
sequence'' and ``blue cloud'' in the CMD of Figure~\ref{fig:cmd}. Motivated
by the bimodal color distribution of galaxies (e.g.,
\citealt{Strateva01,Baldry04}), we construct the division between red and
blue galaxies by fitting the galaxy color distribution with two Gaussian
distributions. Figure~\ref{fig:histriz} shows the probability distribution
function of $r-i$ color (k+e corrected to z=0.55) in six redshift slices as
in Figure~\ref{fig:cmd}. The distribution at each redshift is computed from
galaxy samples that are as complete in luminosity as possible, i.e., we only
use galaxies more luminous than the luminosity given by the intersection of
line (2) and line (3) in Figure~\ref{fig:cmd}, which corresponds to an
$i$-band apparent magnitude of $i_{\rm cmod}=19.46$. The CMASS sample shows a
clear bimodal distribution in color, similar to the findings in SDSS-I/II
\citep{Strateva01,Baldry04,Skibba09a}. We can use the intersection of the
best-fit two Gaussian distributions to divide galaxies into blue and red
samples.

We do not perform the two-Gaussian fit to the galaxy distribution in redshift
interval $0.4<z<0.45$, as these galaxy samples are far from complete. The
blue galaxies are essentially missing from this sample, and even the red
galaxy colors at this redshift are not well described by a Gaussian
distribution. As shown in Figure~\ref{fig:cmd}, galaxies at this redshift
suffer from the $d_\perp$ selection cut, which eliminates the blue galaxies
and a fraction of red ones. In the $0.45<z<0.5$ redshift bin, we still miss
galaxies with $r-i<0.9$, and the distribution is dominated by the Gaussian
profile from the red population. At $0.5<z<0.55$, luminous blue galaxies are
excluded from the sample by the $d_\perp$ cut (see Figure~\ref{fig:cmd}), and
the contribution to the blue Gaussian profile is mainly from faint blue
galaxies. Therefore, at $z<0.55$, blue galaxies in the CMASS sample are far
from complete. For red galaxies, we find that the centers of the color
distribution do not significantly change with redshift. Thus for the analysis
of the whole CMASS sample, we use the redshift-independent color cut for
simplicity (see Equation 7).

The red/blue color division cut shows a mild redshift dependence, becoming
bluer at higher redshift. Since both the color and magnitude used in this
paper have been k+e corrected (i.e., the evolution effects are removed), such
a mild evolution might indicate that the global evolution correction is not
accurate. On the other hand, the photometric errors increase for larger
redshift (see below), making the two Gaussian profiles broader, which can
lead to a shift of the red-blue division cut towards the blue end even if
there is no change in the blue and red populations. Moreover, the blue sample
is generally incomplete due to the selection effects, which may also
introduce additional change of the color cut. Therefore, the weak dependence
of the red-blue division cut on redshift may not reveal much about the
evolution.

\begin{figure*}
\epsscale{1.0}\plotone{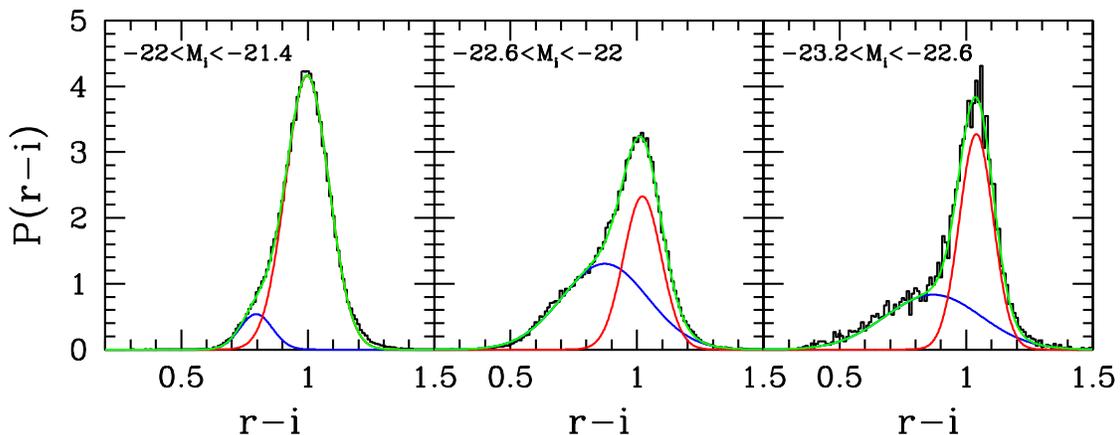}\caption{Similar to
Figure~\ref{fig:histriz}, but for different luminosity intervals at
$0.43<z<0.7$. Note that blue galaxies in the left-most panel are not
complete. (A color version of this figure is available in the online
journal.)} \label{fig:histrimag}
\end{figure*}
We further investigate the dependence of the color distribution on luminosity
for the whole CMASS sample at $0.43<z<0.7$. As shown in
Figure~\ref{fig:histrimag}, the peak of the red sequence, as well as the
intersection of the two Gaussian profiles, become slightly redder as the
luminosity increases, reflecting the well-known tilt of the red sequence. The
tilted red sequence is likely a reflection of differences in the chemical
composition, where the more luminous galaxies are richer in metals while the
smaller galaxies suffer from the loss of metal-enriched gas
\citep{Kodama97,Gallazzi06}. The tilted red sequence may also reflect the
role of dry mergers (i.e., of gas-poor galaxies) in the evolution of red
galaxies \citep[e.g.,][]{Skelton09}, with galaxies increasing their mass
(luminosity) from mergers and becoming older (redder) as a result of stellar
evolution (see \citealt{Faber07} for a comprehensive review). Note that in
the left-most panel, blue galaxies are not complete in the redshift range of
$0.43<z<0.7$ (see Figures~\ref{fig:cmd} and \ref{fig:lumcuts}), which leads
to the non-monotonic behavior across the three panels.

In order to find a reasonable color cut for red and blue galaxies in CMASS,
we fit the bimodal color distribution as a function of luminosity for
galaxies in the range of $0.5<z<0.7$, where the samples are less affected by
incompleteness. The resulting color cut is the one already presented in
Equation~\ref{eq:colorcut}. With such a color cut, if we naively count the
CMASS galaxies disregarding the incompleteness at lower redshifts, we find
that only about $13\%$ of the galaxies in CMASS are blue galaxies, and $80\%$
of these blue galaxies are from $z>0.55$. We emphasize that our color cut is
based on the k+e corrected colors.

\cite{Masters11} proposed an observer-frame color cut of $g-i=2.35$ for CMASS
galaxies, motivated by the color and morphology distribution for a matched
sample between CMASS and Hubble Space Telescope imaging of the Cosmic
Evolution Survey \citep[COSMOS;][]{Scoville07}. They demonstrated that the
$g-i$ cut can be used to separate the elliptical and spiral galaxies in
CMASS. However, there is still some fraction of LRG progenitors with
$g-i<2.35$ due to small amounts of star formation \citep{Tojeiro12b}. For our
purpose of studying the color dependence of clustering, an observer-frame
color cut would mix galaxies from different populations, since galaxies of
the same intrinsic colors will appear to have different observed colors at
different redshifts. In addition, the $g$-band magnitude in CMASS is
especially faint and thus large measurement errors will make $g$-band based
colors prone to spurious fluctuations. We therefore prefer to use the
``intrinsic'' (i.e., k+e corrected) $r-i$ color in our study. We note that
the $z=0.55$ observed $r$- and $i$-bands are close to rest-frame $g$- and
$r$-bands, respectively. So our analysis based on $z=0.55$ $M_i$ and $r-i$
approximates that of rest-frame $M_r$ and $g-r$, as adopted in Z11 for $z\sim
0$ SDSS galaxies.

\begin{figure*}
\epsscale{1.0}\plotone{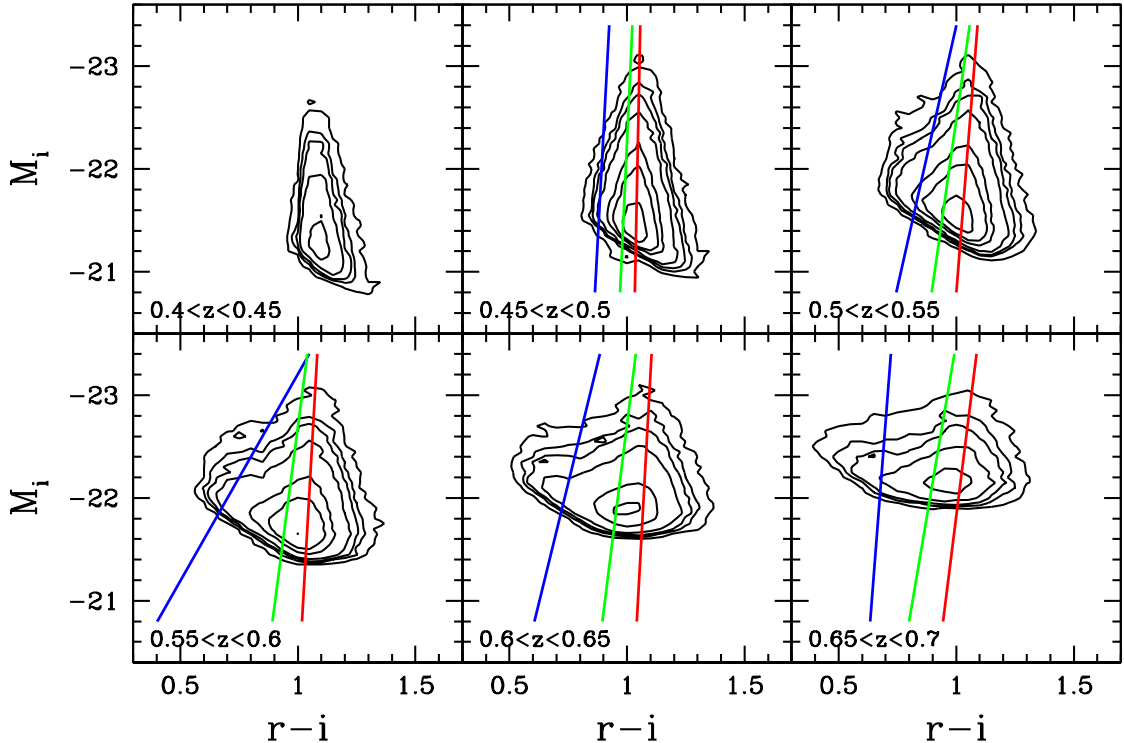}\caption{Our adopted color cuts of
Table~\ref{tab:colorcuts} in the color-magnitude diagram. The \emph{bc},
\emph{br}, and \emph{rr} cuts are shown in the solid lines of colors in blue,
green, and red, respectively. We ignore the redshift interval of $0.4<z<0.45$
due to its high sample incompleteness. (A color version of this figure is
available in the online journal.)} \label{fig:colorcuts}
\end{figure*}

\begin{deluxetable*}{lrrrrrrrrrrc}
\tablewidth{0pt} \tablecaption{\label{tab:colorcuts}luminosity dependent
$r-i$ color cuts for different redshift intervals} \tablehead{\colhead{$z$
range} & \colhead{$a_1$} & \colhead{$b_1$} & \colhead{$a_2$} &
\colhead{$b_2$} & \colhead{$a_3$} & \colhead{$b_3$} &
\colhead{$N_{\rm{blue}}$} & \colhead{$N_{\rm{green}}$} &
\colhead{$N_{\rm{redseq}}$} & \colhead{$N_{\rm{reddest}}$} & \colhead{$V_z
[{\rm Gpc/h}]^3$}}  \startdata
0.45,\ 0.50&-0.023&0.386&-0.0199&0.557&-0.0089&0.849 & 934 & 13233 & 20047 & 25094 & 0.19  \\
0.50,\ 0.55&-0.099&-1.315&-0.0622&-0.398&-0.0346&0.281 & 6215 & 17415 & 24721 & 27361 & 0.22 \\
0.55,\ 0.60&-0.249&-4.777&-0.057&-0.294&-0.0251&0.496 & 2841 & 20323 & 19520 & 18468 & 0.25 \\
0.60,\ 0.65&-0.107&-1.619&-0.054&-0.227&-0.0242&0.538 & 3892 & 13349 & 12261 & 9374 & 0.28 \\
0.65,\ 0.70&-0.034&-0.074&-0.0745&-0.750&-0.0547&-0.193 & 2517 & 6517 & 5797
& 5236 & 0.31
\enddata
\end{deluxetable*}
To study the color dependence in different luminosity and redshift intervals
in more detail, we further decompose the sample into finer color subsamples.
The Gaussian fittings provide the centers and $1\sigma$ widths of the blue
cloud and red sequence, which are used in defining the fine color cuts
\begin{eqnarray}
(r-i)_{\rm{bc}}\ &=&\ \rm{blue\ center}\\
(r-i)_{\rm{br}}\ &=&\ \rm{red\ center}-0.5\times\rm{red\ width}\\
(r-i)_{\rm{rr\ }}\ &=&\ \rm{red\ center}+0.5\times\rm{red\ width}
\end{eqnarray}
With the three cuts, we can form \emph{blue} (below the $bc$ cut),
\emph{green} (between the $bc$ and $br$ cuts), \emph{redseq} (between the
$br$ and $rr$ cuts), and \emph{reddest} (above the $rr$ cut) samples. In each
redshift interval, the luminosity-dependent color cuts are fitted with a
straight line,
\begin{equation}
r-i=a_jM_i+b_j
\end{equation}
where $j=1$, $2$, $3$ for the \emph{bc}, \emph{br}, and \emph{rr} cuts,
respectively. The linear fits for these cuts are listed in
Table~\ref{tab:colorcuts}. For clarity, we show again the CMD in
Figure~\ref{fig:colorcuts}, with the fine color cuts superimposed. The
redshift interval of $0.4<z<0.45$ is omitted because of the high sample
incompleteness.

\begin{figure}
\epsscale{1.0}\plotone{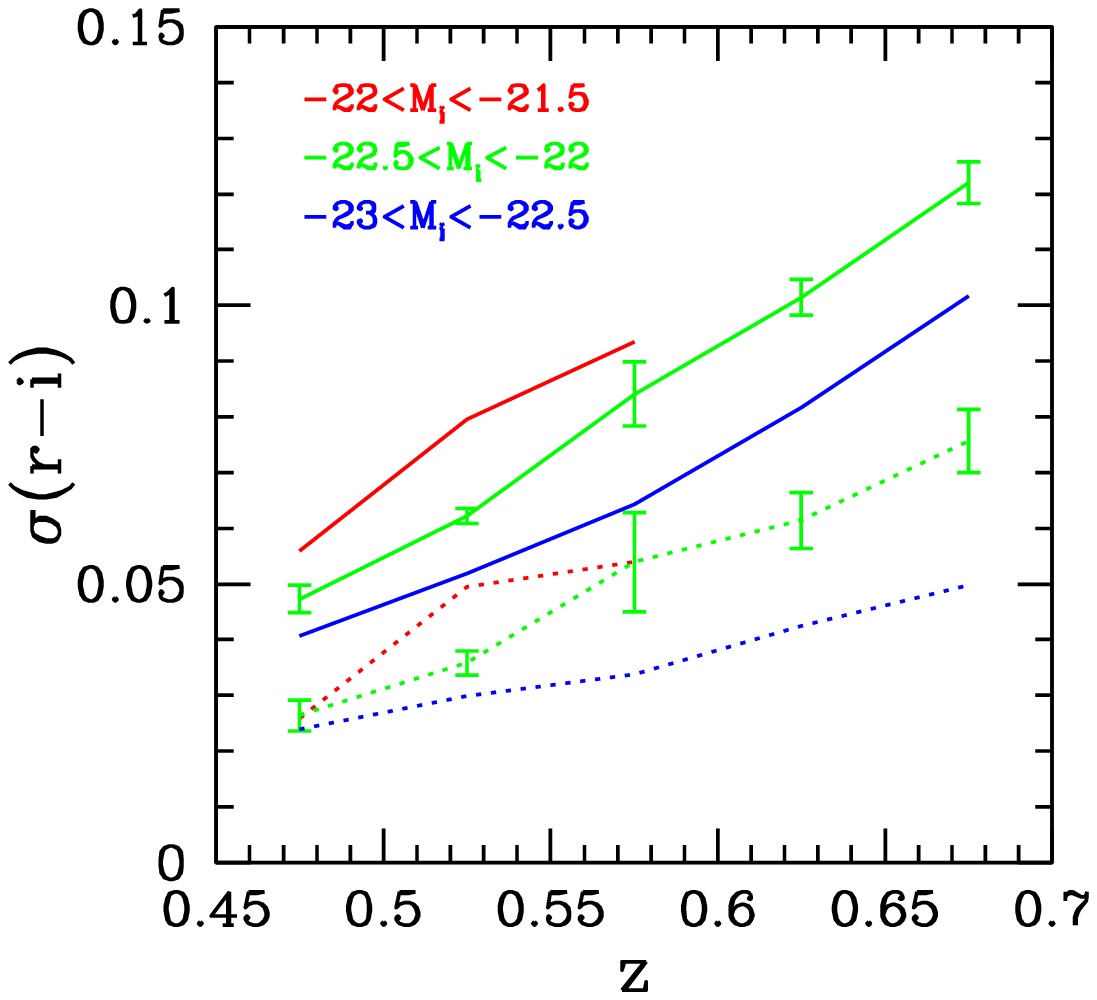}\caption{$1\sigma$ scatter (width) of the red
sequence galaxies as a function of redshift for different magnitude intervals.
The solid lines are the measured scatter, and the dotted lines are the intrinsic scatter by
excluding the photometric errors. The errors are only shown for one luminosity bin for clarity.
(A color version of this figure is available in the online journal.)} \label{fig:rseqwidth}
\end{figure}

Examining Figures~\ref{fig:histriz}-\ref{fig:colorcuts}, there is a
noticeable trend that the width of the red sequence appears narrower for more
luminous galaxies and at lower redshifts. We highlight this effect in
Figure~\ref{fig:rseqwidth}, which presents the 1$\sigma$ width of the red
sequence as a function of luminosity and redshift. Since the photometric
errors become larger for fainter galaxies and at higher redshifts, we
subtract their contribution in quadrature to obtain the intrinsic color width
of the red sequence galaxies. The $r-i$ color photometric errors are
estimated by simply combining in quadrature the errors in $r$- and $i$-band
magnitudes (neglecting any additional errors in the k+e corrections and in
the correlation between $r$- and $i$-band photometric errors). As shown in
Figure~\ref{fig:rseqwidth}, the above trend persists for the intrinsic color
scatter, suggesting an evolutionary effect which we discuss further in
Appendix~\ref{app:models}.

\subsubsection{The Dependence of Galaxy 2PCF on Color}
\label{subsubsec:wpcolor}

\begin{figure*}
\epsscale{1.0}\plotone{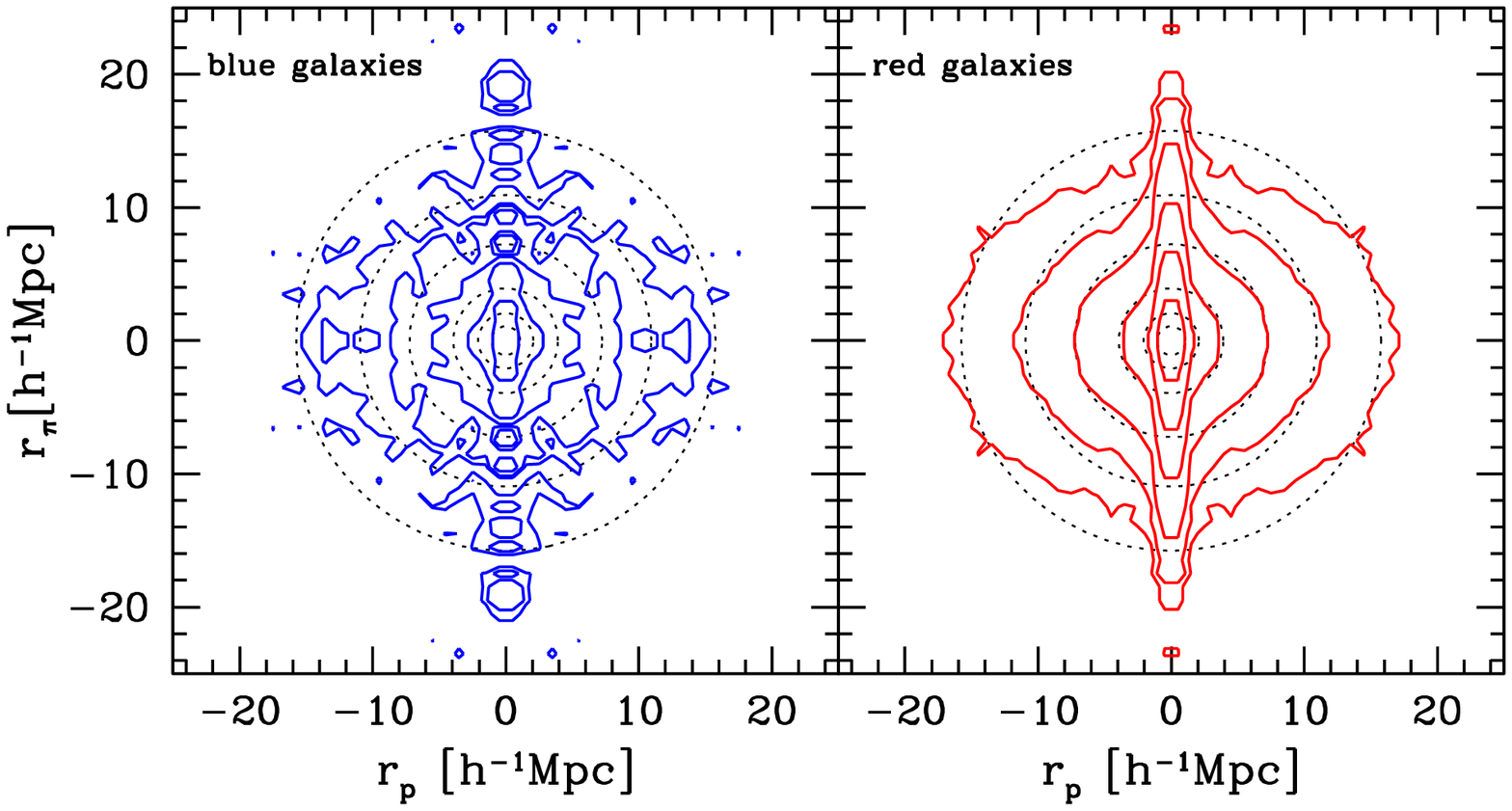}\caption{Measurements of the 3D 2PCF
$\xi(r_p,r_\pi)$ for the blue (left panel) and red (right panel) galaxies in
the whole CMASS sample, defined using the color cut in
Equation~\ref{eq:colorcut}. Contour levels shown are $\xi(r_p,r_\pi)=[0.5, 1,
2, 5, 10, 20]$. The dotted circles in both panels are the angle-averaged
redshift-space correlation function, $\xi(s)$, of the whole CMASS sample for
the same contour values. (A color version of this figure is available in the
online journal.) } \label{fig:xiprb}
\end{figure*}
With the color cuts defined in the previous subsection, we investigate the
dependence of galaxy 2PCFs on the $r-i$ color. First, we examine the 2PCFs
for blue and red galaxies in the whole CMASS sample. The red and blue samples
here are defined using the color cut in Equation~\ref{eq:colorcut}. The
samples are flux-limited (in addition to other selection cuts) and are by no
means complete. The purpose of this exercise is simply to have an overall
view of the difference in red and blue galaxy clustering. The
$\xi(r_p,r_\pi)$ measurements for blue and red galaxies are shown in
Figure~\ref{fig:xiprb}. For reference, the dotted circles in both panels are
the angle-averaged redshift-space correlation function $\xi(s)$. Red galaxies
are more strongly clustered. The ``Fingers-of-God'' feature \citep{Jackson72}
on small scales caused by random motions of galaxies in virialized structures
can be clearly seen for both red and blue galaxies. Red galaxies have a
stronger ``Fingers-of-God'' effect, reflecting their stronger motions within
halos. On large scales (e.g., above $r_p=10\mpchi$, the outmost contours),
the contours for both blue and red galaxies show the flattening trend caused
by coherent large-scale infall \citep{Kaiser87}. On these scales, the Kaiser
squashing effect appears to be stronger for blue galaxies, since the effect
is determined by $\approx \Omega_m^{0.55}/b$ and blue galaxies have a smaller
galaxy bias factor $b$.

\begin{figure*}
\epsscale{1.0}\plotone{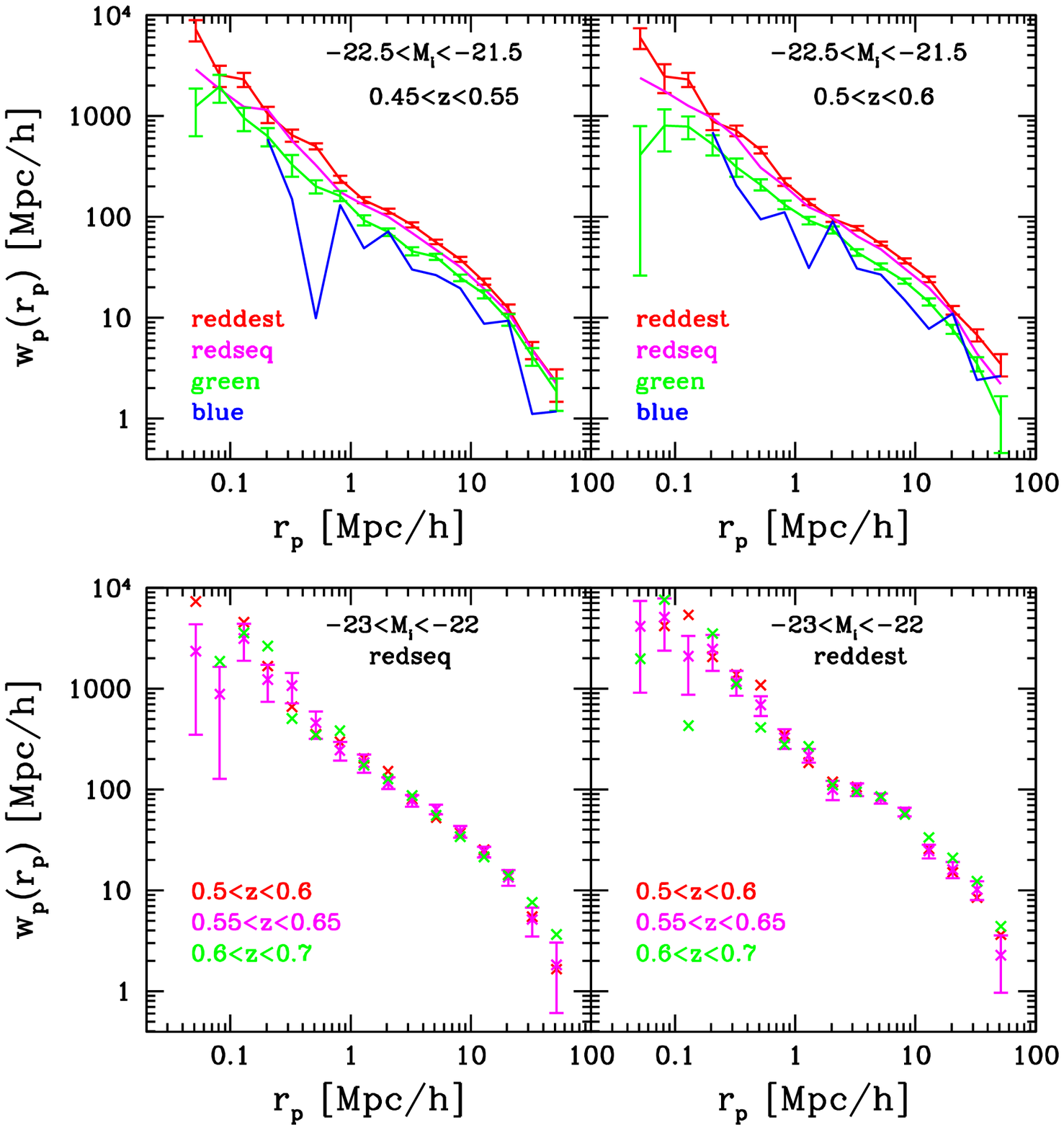}\caption{Color dependence of the
projected 2PCF $w_p(r_p)$. Top panels display the color dependence for the
$-22.5<M_i<-21.5$ sample at two different redshift intervals. The redshift
evolution of $w_p(r_p)$ for the $\emph{redseq}$ and $\emph{reddest}$
subsamples of $-23<M_i<-22$ galaxies is shown on the bottom. (A color version
of this figure is available in the online journal.)} \label{fig:wpcolor}
\end{figure*}
We now investigate the color-dependent 2PCFs as a function of luminosity and
redshift from the fine color samples. In order to minimize the effect of
incompleteness, the luminosity and redshift bins are selected using
Figure~\ref{fig:cmd} to make sure that red galaxies are not affected by the
selection cuts. The blue galaxies are generally not complete at most
redshifts and the results of the $\rm{blue}$ samples need to be interpreted
with care. Nevertheless, the blue samples are still useful in comparison with
the red galaxies.

The main results of the color-dependent 2PCFs are summarized in
Figure~\ref{fig:wpcolor}. The top panels display the dependence of $w_p$ on
color in the magnitude range $-22.5<M_i<-21.5$ at two different redshift
intervals. The trend with color is obvious at both redshifts -- there is a
continuous increase in the clustering amplitude as galaxy color goes from
blue to red. This result is consistent with the behavior observed in the
SDSS-I/II main galaxy sample (Z11). On small scales (below the inflection
scale of 1--2$\mpchi$), there appears to be a trend that redder galaxies have
a steeper slope in $w_p$, which is weaker than that measured by Z11.
According to the HOD modeling result in Z11, for galaxies in a fixed
luminosity range, redder galaxies generally have a higher fraction of
satellites residing in massive halos. Our results therefore implies that a
larger fraction of redder galaxies reside in more massive halos, giving rise
to a larger clustering amplitude. The steepening of $w_p$ on smaller scales
may also indicate a halo mass scale shift with color, leading to a relative
increase in the contribution from the 1-halo central-satellite pairs with
respect to the 1-halo satellite-satellite pairs (see the Appendix A of
\citealt{Zheng09}).

The bottom panels of Figure~\ref{fig:wpcolor} present the 2PCFs for more
luminous red galaxies, with $-23<M_i<-22$, showing the results for the
\emph{redseq} and \emph{reddest} color subsets. The number densities of these
two subsets have a roughly constant value of $0.21\times10^{-4} h^3{\rm
Mpc}^{-3}$ and $0.18\times10^{-4} h^3{\rm Mpc}^{-3}$ over the redshift range,
respectively. There is a stronger inflection in the slope of $w_p$ below
$2\mpchi$ for these luminous galaxies and a trend of steeper slope for the
\emph{reddest} sample than for the \emph{redseq} sample, indicating the shift
in the halo mass scale. The two bottom panels also compare the redshift
evolution for galaxies of the same color and luminosity. No strong evolution
in $w_p$ is found, consistent with the bottom right panel of
Figure~\ref{fig:wplum}. The implications of these results are discussed
further in Section~\ref{sec:conclusion}.

\subsection{Red Galaxy Samples with Fixed Number Density}
\label{subsec:fixn}
\begin{figure*}
\epsscale{1.0}\plotone{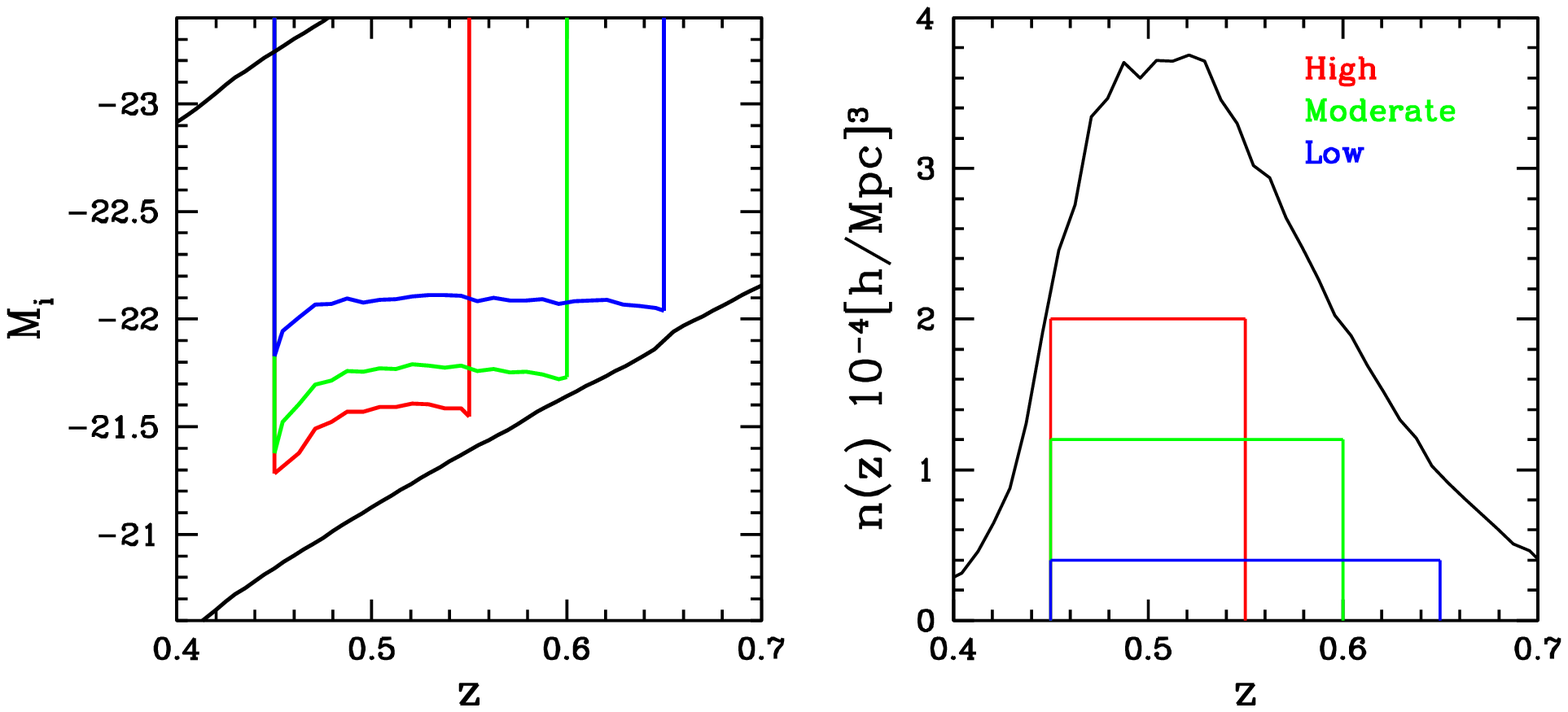}\caption{Construction of fixed number density
samples for the red galaxies, as defined by Equation~\ref{eq:colorcut},
corresponding to low, moderate, and high $n(z)$.  The galaxies in each sample
are selected by a redshift-dependent luminosity threshold $M_i(z)$, shown in
the left panel for the three samples by the blue, green and red lines,
respectively. (The two black lines delineate the $i$-band flux limits as in
the right panel of Figure~\ref{fig:lumcuts}.) The right panel shows the
corresponding $n(z)$ for the three samples, while the black solid line is the
overall number density distribution of CMASS galaxies. (A color version of
this figure is available in the online journal.)} \label{fig:fixn}
\end{figure*}
In previous sections, we constructed galaxy samples in certain luminosity and
color bins, in an attempt to minimize the influence of incompleteness caused
by target selection cuts. Motivated by a simple passive evolution model, we
can further define galaxy samples with fixed number density at different
redshifts \citep{White07,Brown08,Wake08}. If during the evolution each galaxy
in a sample retains its identity, experiencing no merger or disruption, the
number density of the galaxy sample would not change with redshift. The
evolution of the 2PCF of such a galaxy sample can be readily predicted
\citep{Fry96}. If, in addition, no star formation occurs in these galaxies
during the process, their stellar population would evolve passively and can
be readily modeled \citep{Wake08,Tojeiro12b}. Comparing to such predictions
allows a rough determination of the extent of evolution in the red galaxy
samples.

We construct three such samples for the red galaxies, which may be expected
to resemble a passively evolving population, with fixed low, moderate and
high number densities. For each sample, the fixed number density is achieved
by finding a (redshift-dependent) luminosity threshold $M_i(z)$ and selecting
all galaxies with luminosity above this threshold, as shown in
Figure~\ref{fig:fixn}. The luminosity and redshift ranges are chosen to
reduce the sample incompleteness caused by the selection cuts. The low,
moderate, and high number density samples have $n(z)$ of $0.4\times10^{-4}$,
$1.2\times10^{-4}$, and $2\times10^{-4}\, h^3{\rm Mpc}^{-3}$, respectively.

As seen in Figure~\ref{fig:fixn}, the fixed number density thresholds
correspond globally to rough luminosity thresholds, decreasing with
increasing number density as expected.  The luminosity thresholds $M_i(z)$
stay roughly constant for the red galaxies at $z>0.48$. Since the
luminosities in our study have been k+e corrected, this result implies that
the stellar population in these red galaxies evolves passively. The drop
below $z<0.48$ is likely caused by the incompleteness of galaxies at lower
redshift due to the CMASS selection cuts, as discussed in
Section~\ref{subsec:incomp}.

\begin{figure*}
\epsscale{1.0}\plotone{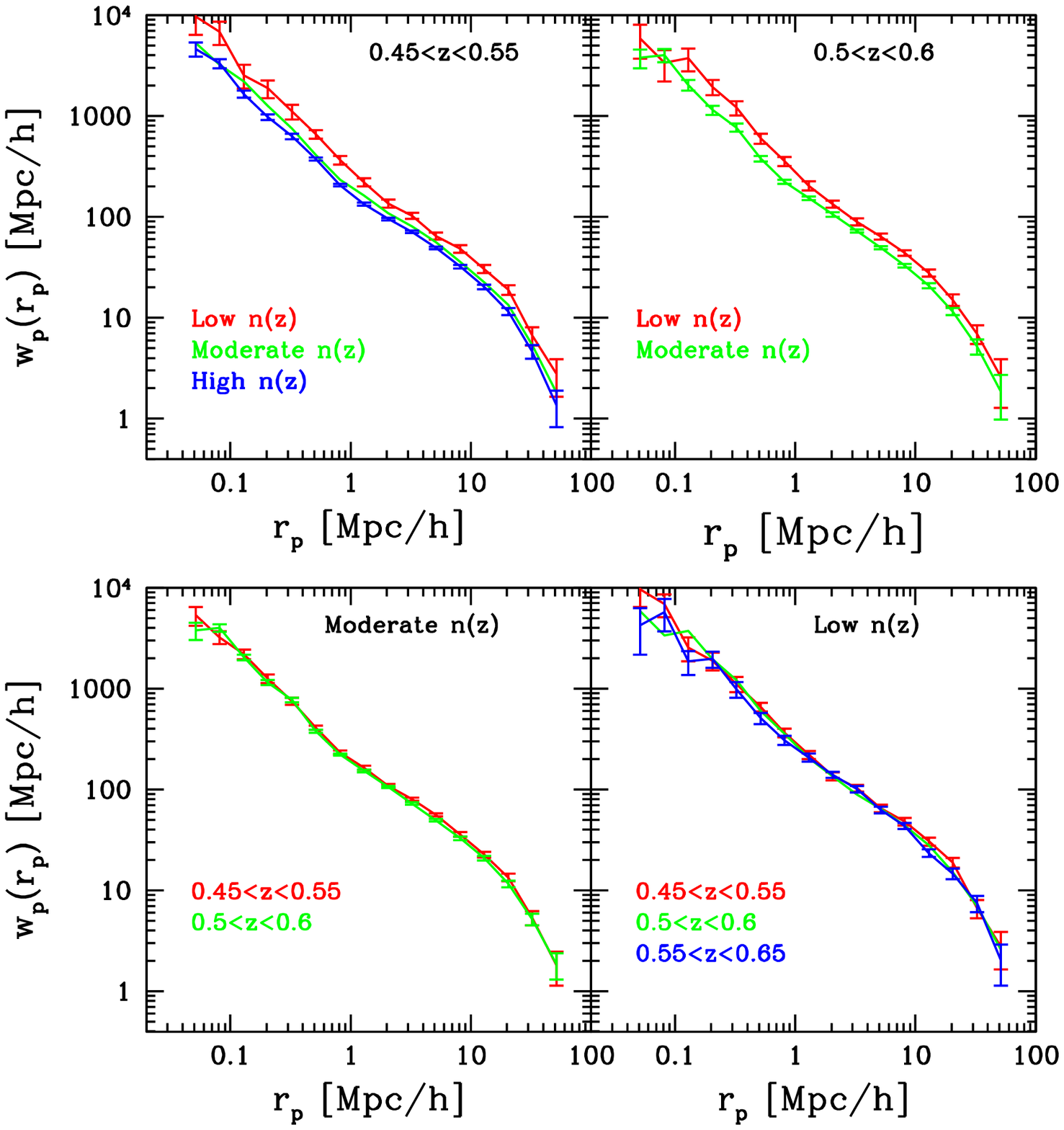}\caption{Projected 2PCFs, $w_p(r_p)$, for
fixed number density samples of CMASS red galaxies. The top panels show the
dependence on number density for two redshift bins, and the bottom panels
show the redshift dependence for the moderate and low number density samples.
(A color version of this figure is available in the online journal.)}
\label{fig:wpfixn}
\end{figure*}
The projected 2PCFs for the fixed number density samples of red galaxies are
presented in Figure~\ref{fig:wpfixn}. The top panels show that the clustering
strength is inversely proportional to the number density, such that galaxies
in the lower number density samples are more clustered. These results are
generally consistent with the luminosity dependence discussed in
Section~\ref{subsec:lum}, as the different values of $n(z)$ effectively act
as different luminosity thresholds. The shapes of the 2PCFs for different
$n(z)$ samples are similar. The bottom panels compare the 2PCFs of the same
$n(z)$ samples at different redshift intervals. We find that the 2PCFs at
different redshifts have similar clustering strength, again consistent with
the results for the luminosity dependence of red galaxies
(Figure~\ref{fig:wplum}, bottom-right panel). We discuss the implications of
these results next.

\subsection{Galaxy Bias}\label{subsec:bias}
With the measured galaxy 2PCFs and the theoretical matter 2PCF for the
cosmology adopted in this paper, we can infer the galaxy linear bias factor
$b$ from the square root of the ratio between the galaxy and dark matter
2PCFs. The evolution of the linear bias factor can provide hints about the
evolution of galaxy samples. In this subsection, we present the results on
galaxy linear bias factor $b$ as a function of galaxy luminosity, color,
number density, and redshift, and discuss the possible implications on galaxy
evolution. Since blue galaxies in the CMASS sample are generally far from
complete, we will focus our discussion on the bias evolution of the red
galaxies.

At each redshift, for each galaxy sample, we estimate the linear bias factor
$b$ by taking the square root of the ratio between the measured projected
galaxy 2PCF and the non-linear dark matter projected 2PCF computed at the
corresponding redshift, where the latter is calculated using a modified {\tt
halofit} model \citep{Smith03} with the \citet{Eisenstein98} power spectrum
parameterization. More specifically, we fit the ratio of galaxy and matter
$w_p(r_p)$ with a single parameter $b$ on scales $3\mpchi<r_p<25\mpchi$,
using the full covariance matrix of the galaxy $w_p(r_p)$.

\citet{Fry96} shows that the passive evolution prediction for the linear bias
factor $b(z)$ follows
\begin{equation}
\label{eq:fry}
b(z)=1+\frac{b_0-1}{D(z)},
\end{equation}
where $D(z)$ is the linear growth factor at redshift $z$ and $b_0$ is the
bias factor at $z=0$ ($D(0)=1$). From this relation, the redshift evolution
of galaxy 2PCF for the passively-evolving population can be expressed as
\begin{equation}
\xi(z)=[b(z)D(z)]^2\xi_m(0)=[D(z)+(b_0-1)]^2\xi_m(0),
\end{equation}
where $\xi_m(0)$ is the matter 2PCF at $z=0$. Here we use the word
``passive'' to mean that during the evolution, each galaxy in the sample
keeps its identity and there is no merger or disruption that changes the
population. For the CMASS sample considered in this paper, the galaxy bias
factor is usually greater than unity. Therefore, according to the above two
equations, for passively-evolving galaxies, we expect that with decreasing
redshift  the amplitude of 2PCF increases while the bias factor decreases.

\begin{figure*}[htbp]
\epsscale{1.0}\plotone{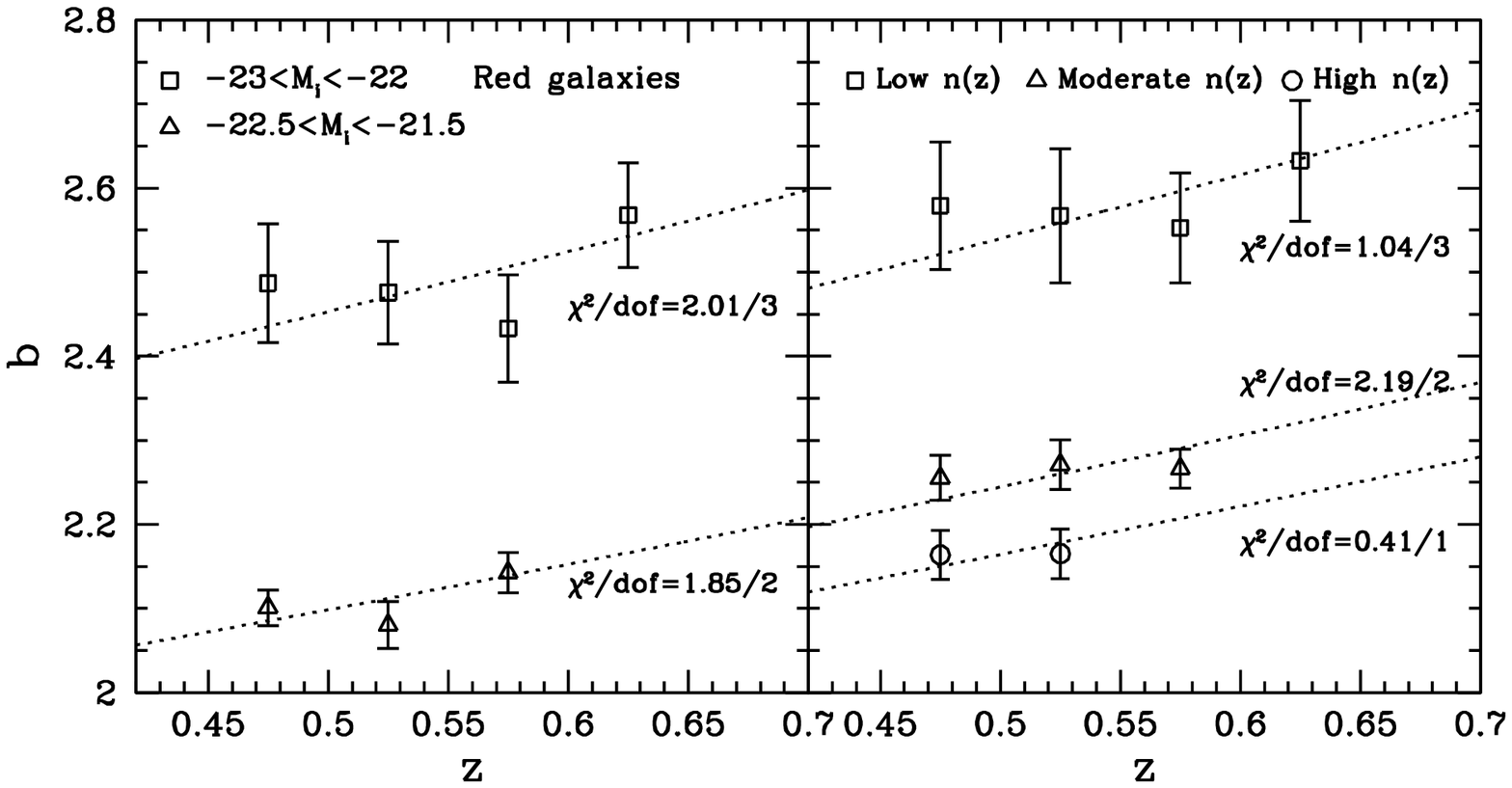}\caption{Linear bias factor, $b$, as a
function of redshift for the CMASS red galaxies in different luminosity
samples (left panel) and different fixed $n(z)$ samples (right). The symbols
represent the measured galaxy bias $b(z)$ by fitting $w_p(r_p)$ over
$3\mpchi<r_p<25\mpchi$ relative to the theoretically predicted one for dark
matter. The dotted lines are the best-fit passive evolution predictions by
fitting $b(z)$ using the \citet{Fry96} relation. The goodness of fit
$\chi^2/dof$ is also given for each set of samples.} \label{fig:passive}
\end{figure*}
Figure~\ref{fig:passive} shows the bias redshift evolution for luminosity bin
samples (left) and for the fixed number density samples (right). The bias
factors are measured for non-overlapping redshift bins. The fitted bias
factors support our results in the previous subsections that more luminous
(or lower density) samples are more strongly clustered. The dotted lines in
Figure~\ref{fig:passive} are the best-fit \citet{Fry96} relation for each
sample, with $b_0$ as the single fitting parameter. For the two
luminosity-bin red galaxy samples, we find that they roughly follow the
passive evolution prediction. Strictly speaking, each luminosity bin sample
does not conserve the number density at different redshifts. So by definition
it is not a passively-evolving population. However, these number density
differences may be accounted for by slightly changing the luminosity
thresholds of the luminosity bin sample at each redshift (e.g., as a result
of an imperfect k+e correction). The measured bias would not be sensitive to
such an adjustment. In such a sense, comparing their bias evolution to the
Fry relation can still be meaningful.

For both the luminous and faint samples, there is suggestive evidence that
the clustering at intermediate redshifts ($z=0.575$ for the luminous sample,
and $z=0.525$ for the faint sample) is slightly weaker than that from the
best-fit passive evolution. Similar deviations from passive evolution were
found in other works (e.g., \citealt{White07,Wake08,Sawangwit11}). However,
the large measurement errors in our results make the deviation only at about
$1\sigma$ level, limiting our ability for a solid conclusion.

If this deviation is robustly established, with more accurate future
measurements using CMASS data over a larger survey area, it would imply a
significant contribution from processes that break passive evolution, such as
feedback from active galactic nuclei shutting off star formation, disruption
of satellites in massive halos, and mergers of galaxies
\citep{Bell04,Faber07,Skelton09}. The signature may be related to the overall
migration of blue galaxies to the red sequence \citep{Martin07}, in which
case, it would indicate appreciable migration by $z{\sim}0.55-0.6$,
consistent with the prediction of \cite{Faber07}. A sophisticated model is
needed to disentangle the contributions from the different evolutionary
processes.

The more reliable samples to study the passive evolution are the ones with
fixed number densities, as described in Section~\ref{subsec:fixn}. The
results for our three samples are shown in the right panel of
Figure~\ref{fig:passive}. The low $n(z)$ sample appears consistent with
passive evolution in the redshift range $0.45<z<0.65$, within the (large)
error bars on the measurements. This behavior is similar to the $-23<M_i<-22$
sample. In fact the low $n(z)$ sample is close to a luminosity-threshold
sample of $M_i<-22$, as shown in Figure~\ref{fig:fixn}. For the two samples
with higher $n(z)$, their bias evolution is consistent with passive evolution
for the smaller redshift ranges probed, which is in agreement with the
conclusions of \cite{Tojeiro12b}. Within the current uncertainties, however,
it is not possible to make strong statements regarding confirming or ruling
out passive evolution. We will revisit this with more accurate measurements
with future larger CMASS samples.

\begin{figure*}
\epsscale{1.0}\plotone{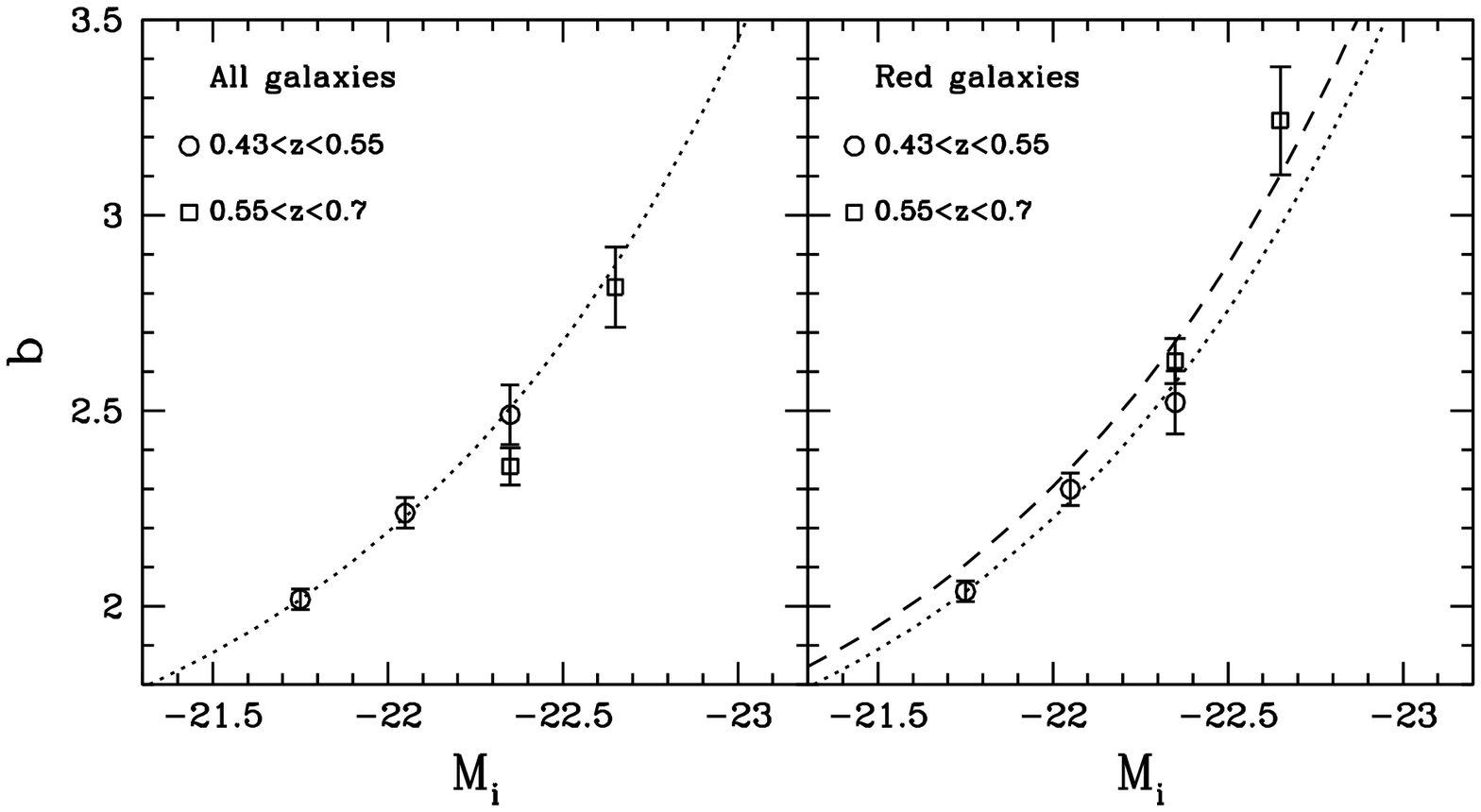}\caption{Linear bias factor, $b$, as a
function of luminosity in two redshift ranges for all CMASS galaxies (left)
and for the red galaxies only (right). The dotted curves are the
bias--luminosity relation, Equation~\ref{eq:biasfit}, with $c_2=0.33$ for all
galaxies and $c_2=0.35$ for the red galaxies, for the lower redshift range.
The dashed curve in the right panel is the low-redshift relation shifted to
the higher-redshift range according to the passive evolution prediction. }
\label{fig:biaslum}
\end{figure*}
Finally, we present the dependence of the bias factor on luminosity in
Figure~\ref{fig:biaslum} for all galaxies and for the red galaxies only.
Generally, more luminous and redder galaxies at higher redshifts have larger
bias factors. For the fainter samples, the bias factors are similar in the
two cases, since at the faint end red galaxies dominate the CMASS sample (the
majority of the faint blue galaxies are excluded by the selection cuts). The
observed dependence of galaxy bias factor on galaxy luminosity is broadly
similar to that for local galaxy samples (e.g., \citealt{Norberg01}, Z11),
but it is non-trivial to compare in detail due to the many differences in
sample selection, redshift, k+e corrections and magnitudes.

We fit the bias-luminosity relation with a commonly-used simple functional
form \citep{Norberg01,Zehavi05b},
\begin{equation}
b/b_p=c_1+c_2L/L_p
\label{eq:biasfit}
\end{equation}
We define $L_p$ as the mean luminosity of galaxies in the faintest luminosity
bin. This sample of galaxies has $b=b_p$, so by construction, $c_1=1-c_2$. We
fit this functional form to the luminosity-dependent bias measurements at
$0.43<z<0.55$, finding $c_2=0.33$ for all galaxies and $c_2=0.35$ for the red
galaxies (shown as the red dotted curves in Figure~\ref{fig:biaslum}). At
higher redshifts, $0.55<z<0.7$, the bias factors for all galaxies show a
decrease compared to the lower-redshift relation due to the inclusion of more
blue galaxies at high redshifts. In contrast, the bias factors for the red
galaxies globally increase with redshift, as expected from passive evolution.
The dashed curve in the right panel shows the low-redshift relation shifted
to high redshift according to the Fry relation prediction, which is in
agreement with our measurements.

\section{Conclusion and Discussion}\label{sec:conclusion}

In this paper, we measure the luminosity and color dependence of the galaxy
2PCFs based on ${\sim}260,000$ BOSS CMASS DR9 galaxies over a ${\sim}
3,300\deg^2$ survey area in the redshift range of $0.43<z<0.7$ and study the
implications on galaxy formation and evolution.

We first measure the 2PCF for the entire sample. If approximated by a
power-law, the 2PCF has a correlation length of $r_0=7.93\pm0.06\mpchi$ and a
slope of $\gamma=1.85\pm0.01$, consistent with the measurements presented by
\citet{White11} and \citet{Nuza12}. We also construct color and luminosity
subsamples. To reduce the influence of sample incompleteness caused by the
target selection criteria of CMASS galaxies, we carefully account for the
selection cuts in the color-luminosity distribution of galaxies at each
redshift interval in defining our subsamples. These subsamples in certain
color and luminosity bins are close to complete and volume-limited. In order
to compare the clustering of galaxy populations at different redshifts, we
perform k+e correction to the magnitudes and colors of the galaxies using
FSPS models. Such corrections are unavoidably dependent on the specific
stellar evolution models used. However, as shown in
Appendix~\ref{app:models}, our general results on the clustering analysis
would not be significantly affected.

We find that for all redshift intervals probed, more luminous galaxies are
more strongly clustered, consistent with previous studies for galaxies at
different redshifts, such as SDSS-I/II main sample galaxies at $z\sim 0.1$
\citep{Zehavi05b,Zehavi11}, SDSS LRG galaxies at $z\sim 0.35$
\citep{Zehavi05a}, and DEEP2 galaxies at $z\sim 1$ \citep{Coil06}. At each
redshift, the large-scale galaxy bias factor of CMASS galaxies shows a linear
dependence on galaxy luminosity, similar to that for lower redshift galaxies
(\citealt{Norberg02}, Z11), but with different coefficients in the
bias-luminosity relation. We divide galaxies globally into a blue and a red
population. For each population, we find a similar clustering trend -- an
increasing clustering strength with luminosity. For blue galaxies, our
results are in line with that of Z11 for SDSS galaxies. For red galaxies, Z11
find that both the most luminous and faintest galaxies exhibit stronger
small-scale ($\lesssim 2 h^{-1}{\rm Mpc}$) clustering than the samples of
intermediate luminosity, which can be explained as a large satellite fraction
in the faintest sample and high mass of host halos for the most luminous
sample. Because CMASS selects mostly luminous galaxies, we are not able to
investigate the trend towards faint red galaxies, but for the luminous red
galaxies, our results agree with that in Z11.

We further investigate the dependence of clustering on galaxy color, using
finer color cuts. For fixed redshift and luminosity, we find that redder
galaxies exhibit stronger clustering, similar to the trend found for the SDSS
main galaxies \citep{Zehavi05b,Zehavi11,Li06}. Interestingly, such a trend
exists even within the red sequence, consistent with the finding of
\citet{Zehavi05b,Zehavi11}. The trend is different from that of DEEP2
galaxies in \citet{Coil08}, where no clear color dependence is seen across
the red sequence. If this difference is caused by galaxy evolution, it
implies that the color dependence in the red sequence emerges during the
redshift range of $0.7<z<1.0$. The emergence of the dependence may signal the
contribution of substantial amounts of mergers and inflow of blue galaxies to
the buildup of the red sequence.

We also construct subsamples of red galaxies with fixed number densities by
applying redshift-dependent luminosity thresholds, and compare their
clustering with the theoretical prediction of passively-evolving galaxies
\citep{Fry96}. We find that the evolution of the large-scale galaxy bias
factors for all the three CMASS subsamples considered in this paper are
consistent with that from the Fry relation, within the relatively large
uncertainties in the measured bias factors, which suggests that the red
galaxies in the CMASS sample roughly follow passive evolution from $z=0.7$ to
$0.45$. In contrast, from HOD modeling of clustering of red galaxies in NDWFS
\citep{Brown08}, \citet{White07} found that passive evolution from $z\sim
0.9$ to $z\sim 0.5$ would predict too many satellite galaxies in high-mass
halos and concluded that about one-third of these satellites must have
experienced merging or disruption. The apparent discrepancy between our
result and that in \citet{White07} can be explained by the difference in the
number densities of galaxy samples. The NDWFS samples analyzed in
\cite{White07} have a constant comoving number density of
$n(z)=10^{-3}h^3{\rm Mpc}^{-3}$, about one order of magnitude higher than the
ones we study here. Thus, their constant number density samples include
fainter red galaxies (which have a larger contribution from satellite
galaxies). In contrast, galaxies in our more luminous samples are
predominantly luminous central galaxies that roughly follow passive evolution
(but see also \citealt{Wake08,Sawangwit11}). These results seem to support a
scenario in which mergers and disruption play an important role for the
evolution of low-mass red galaxies.

Our investigation of the color-luminosity distribution at each redshift
reveals two notable trends in the width of the red sequence. The red sequence
becomes narrower towards the high-luminosity end, and it becomes narrower
towards lower redshifts. Similar results are also seen in galaxies at both
lower and higher redshifts \citep[e.g.,][]{Bell04,Skibba09a, Whitaker10}. The
color scatter in the red sequence reflects the distribution of the ages of
stellar population, dust extinction, and metallicity. At a given redshift,
fainter galaxies show a more diverse distribution of these quantities,
leading to a wider distribution in color. Passive evolution makes galaxies
redder and largely reduces the color difference caused by the distribution of
the ages of stellar population, leading to a narrower red sequence towards
lower redshifts.

The inferences in this paper about the evolution of CMASS galaxies from the
measured color and luminosity dependent clustering are still based on simple
clustering models and interpretations. A further, natural step to interpret
these results is to perform HOD modeling of our measurements, which will
allow us to better study galaxy formation and evolution by incorporating
knowledge about the dark matter halo formation and evolution. We expect that
improved measurements from larger BOSS samples in the future and detailed HOD
modeling will greatly advance our understanding of the evolution of massive
galaxies.

\acknowledgments

We thank Joanne Cohn, Peder Norberg, Rom\'an Scoccimarro and Benjamin Weiner
for helpful discussions. We thank the anonymous referee for useful comments.
HG, IZ and ZZ were supported by NSF grant AST-0907947. RAS was supported by
NSF grant AST-1055081 and MECS was supported by NSF grant AST-0901965.

Funding for SDSS-III has been provided by the Alfred P. Sloan Foundation, the
Participating Institutions, the National Science Foundation, and the U.S.
Department of Energy Office of Science. The SDSS-III web site is
http://www.sdss3.org/.

SDSS-III is managed by the Astrophysical Research Consortium for the
Participating Institutions of the SDSS-III Collaboration including the
University of Arizona, the Brazilian Participation Group, Brookhaven National
Laboratory, University of Cambridge, Carnegie Mellon University, University
of Florida, the French Participation Group, the German Participation Group,
Harvard University, the Instituto de Astrofisica de Canarias, the Michigan
State/Notre Dame/JINA Participation Group, Johns Hopkins University, Lawrence
Berkeley National Laboratory, Max Planck Institute for Astrophysics, Max
Planck Institute for Extraterrestrial Physics, New Mexico State University,
New York University, Ohio State University, Pennsylvania State University,
University of Portsmouth, Princeton University, the Spanish Participation
Group, University of Tokyo, University of Utah, Vanderbilt University,
University of Virginia, University of Washington, and Yale University.

\appendix

\section{A. Different Stellar Evolution Models}\label{app:models}
\begin{figure*}
\epsscale{1.0}\plotone{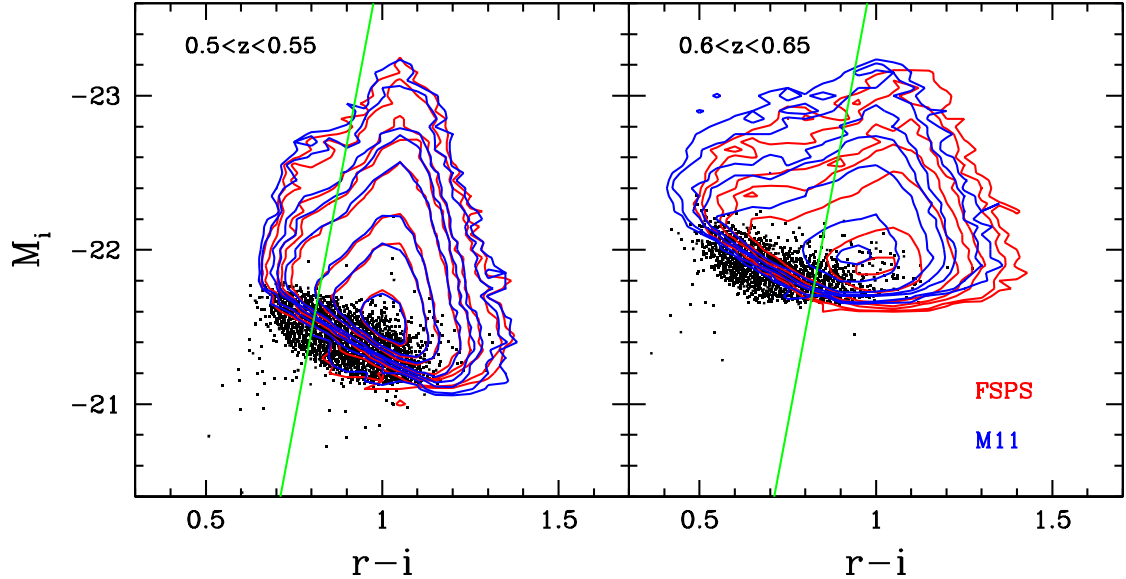}\caption{Color-magnitude diagram for
CMASS galaxies using the FSPS and M11 stellar evolution models, for two
redshift intervals of $0.5<z<0.55$ and $0.6<z<0.65$. The red contour lines
are for the FSPS model and the blue ones are for the M11 model. The green
line is our adopted color cut using the FSPS models. The points are the
galaxies from the CMASS Sparse Sample (see text). (A color version of this
figure is available in the online journal.)}\label{fig:cmd_models}
\end{figure*}

In this paper, we correct for the k+e effects using the FSPS model, as
mentioned in \ref{subsec:incomp}. \cite{Tojeiro12b} compare the FSPS model
and the stellar evolution model of Maraston \& Str\"omb\"ack (2011; M11), and
conclude that both models provide similar star formation histories with
similar mass-weighted ages. \cite{Tojeiro12b} also compare the large-scale
clustering using both models, and find that they produce consistent results.
We show in Figure~\ref{fig:cmd_models} the color-magnitude diagram obtained
using the two different models at two redshift ranges. Contour lines for the
FSPS model are shown in red and for the M11 model in blue. The green line is
our proposed color cut (using the FSPS model). The two models predict similar
k+e corrections for $z<0.55$, but at higher redshifts the M11 model appears
to produce more luminous and bluer galaxies. The overall shapes of the
distributions for the two models, however, are quite similar. In particular,
using alternatively the M11 model would not tilt  the slope of the red
sequence. Using the M11 model, the detailed color cuts would be changed
accordingly but the clustering dependence on the luminosity and color are not
expected to change.

From the CMD, it appears that CMASS galaxies have a tail of faint
extremely-red galaxies, which is different from the properties of the SDSS
main sample (see, e.g., Z11). This appearance, however, is partly caused by
the selection cuts, which remove most of the faint blue galaxies. To clarify
this effect, we show in Figure~\ref{fig:cmd_models} the galaxies from the
BOSS CMASS Sparse Sample, which includes fainter and bluer galaxies by
extending the sliding cut (Equation~\ref{eq:slidecut}) to
\begin{equation}
i_{\rm cmod} < 20.14 + 1.6(d_{\perp} - 0.8)
\end{equation}
\citep{Padmanabhan13}. These faint blue galaxies are sparsely sampled to
yield approximately five objects per square degree. It is clear that the red
sequence has more triangular distribution in CMASS, as a result of the
increase in the width of the color distribution for fainter galaxies
(Figure~\ref{fig:rseqwidth} and discussion thereof). A similar shape of the
red sequence is also observed at higher redshifts \citep[see e.g., Figure~1
of][]{Whitaker10}. The narrow red sequence shown in Z11  is caused by the
fact that their CMD is for the flux-limited main sample galaxies. Many
fainter galaxies at higher redshifts are excluded by the faint flux limit. If
shown for a volume-limited sample, the ``triangle'' shape of the red sequence
is more apparent (see e.g., Figure 2 of \citealt{Skibba09a}). Since this
Sparse Sample does not have the same selection as other CMASS galaxies, we do
not include it in our analysis, but these galaxies can be useful for studying
evolution of blue galaxies that have smaller stellar masses.

\section{B. Jackknife Error Estimates}\label{app:jack}

We use the jackknife resampling method to construct the error covariance
matrices (Equation~\ref{eq:jk}). In principle, it is preferable to derive
covariance matrices from large numbers of realistic mock catalogs, each
matching the observed galaxy properties, survey geometry, and selection
functions. However, our tests below demonstrate that the jackknife error
estimates perform quite well and are sufficient for our purposes. Moreover,
it is a far more practical tool when working with many subsamples of
different clustering properties. We do not currently have available mocks
with suitable modeling of the galaxy luminosity function and with the correct
galaxy distribution on small scales. The large set of mock catalogs used by
\cite{Anderson12} are constructed by populating dark matter halos in
simulations with galaxies according to the HOD model fitted to the
redshift-space 2PCF $\xi(s)$ on large scales ($30\mpchi<s<80\mpchi$). The
measured large-scale 2PCFs of galaxies are reasonably reproduced in these
mocks, while the small-scale 2PCFs are not matched \citep[see][for
details]{Manera13}. Although the small scale 2PCFs measured from these mocks
generally deviates from the real data, the mocks can still be used to
evaluate the validity of the jackknife method, and the appropriate number of
jackknife subsamples to use.

\begin{figure}
\epsscale{1.0}\plotone{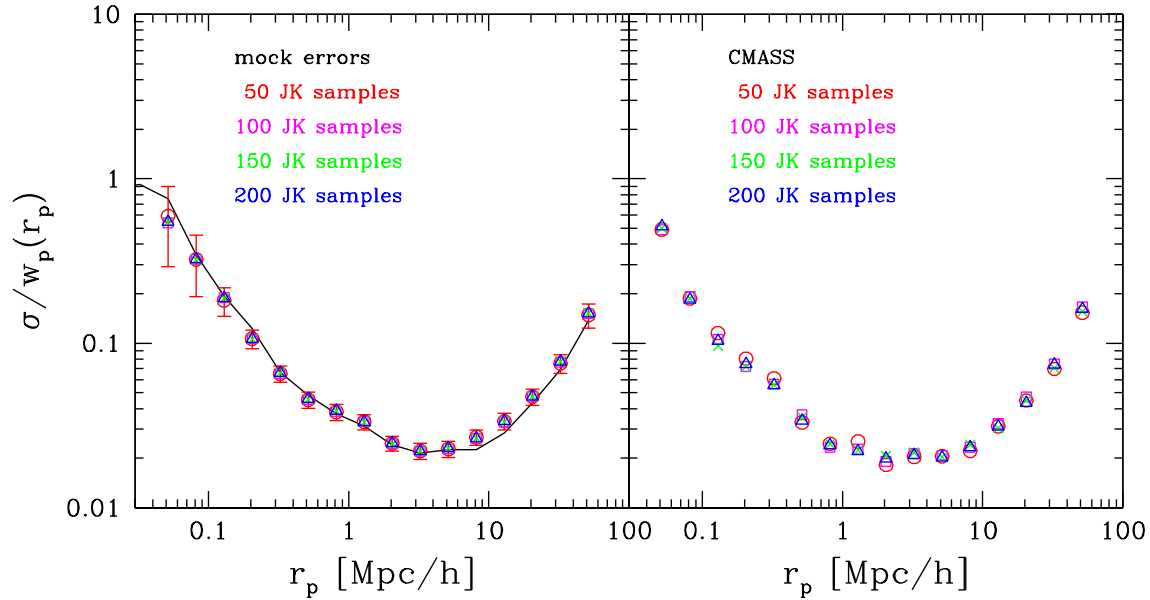}
\caption{Diagonal jackknife errors on the projected 2PCF, $w_p$, estimated
from mock catalogs
(left-hand side) and the real data (right), using different number of
jackknife samples. In the left panel, the solid curve shows the
fractional errors estimated from the variance among $w_p$ measurements in
100 mock catalogs \citep{Manera13}. The symbols
are the average (over the 100 mocks) of the fractional jackknife errors
in each catalog. The errorbars plotted reflect the variation among the jackknife
estimates across the 100 mocks. The number of jackknife samples in each mock
ranges from 50 to 200, as indicated by the
color of the symbols.   On the right we show the fractional jackknife errors
estimated from our ``one realization'' of the actual CMASS data, for
different number of jackknife samples.
(A color version of this figure is available in the online journal.)}
\label{fig:jackknife}
\end{figure}
\begin{figure}
\epsscale{0.6}\plotone{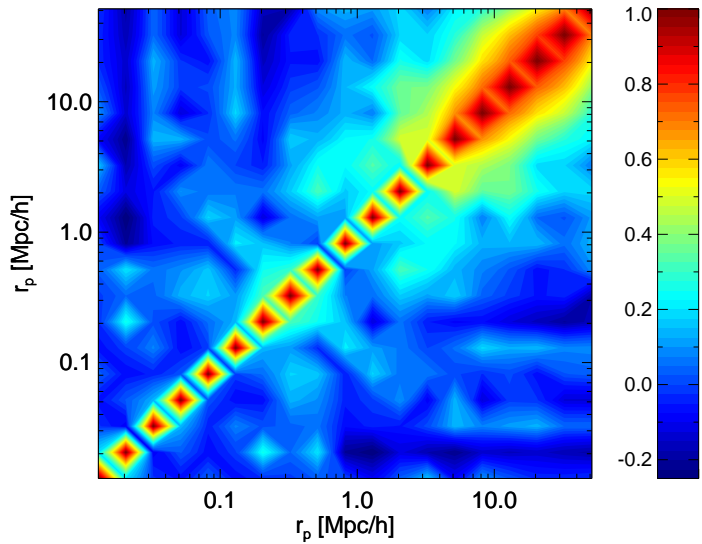} \caption{The covariance matrix of $w_p$ for
the whole CMASS sample, estimated with 100 jackknife samples, normalized by
the diagonal elements of the matrix.  The scale on the right shows the color
scheme, representing the level of covariance on the different scales. (A
color version of this figure is available in the online journal.)}
\label{fig:cov}
\end{figure}
In the left panel of Figure~\ref{fig:jackknife}, we compare the fractional
diagonal errors of $w_p$ from variations among 100 mock catalogs of
\cite{Manera13} (solid curve) and those from applying the jackknife
resampling method to the mocks with different number of jackknife samples
(symbols with different colors). The jackknife error estimates show excellent
consistency with those estimated from variations among mocks, especially for
scales less than a few $\mpchi$. Around $10\mpchi$, the jackknife method
slightly overestimates the errors (by $\sim$15\%). The jackknife errors also
appear to be insensitive to the total number of jackknife samples used, with
a general convergence for 100 or more jackknife samples. We have also checked
the off-diagonal elements in the covariance matrices and find that the
jackknife estimates are generally consistent with the variations-among-mocks
estimates, although there are somewhat larger fluctuations of the jackknife
estimates from mock to mock. All the results are consistent with those found
and discussed in \cite{Zehavi05b}.

\cite{Norberg09} have performed a comprehensive study of a variety of error
estimators for dark matter correlation functions in $N$-body simulations,
comparing estimates such as jackknife and bootstrap to those derived from
multiple independent catalogs. They find good agreement between jackknife and
external estimates for the variance in $w_p(r_p)$ on large scales, consistent
with our results in Figure~\ref{fig:jackknife}. However, \cite{Norberg09}
find that the jackknife method significantly overestimates the errors in
$w_p$ on small scales (e.g., by 40\% at $r_p<1\mpchi$), while we do not see
such a large difference in our results. It is worth noting that we differ in
our implementation of the jackknife method. While \cite{Norberg09} estimate
the jackknife errors by dividing the simulation boxes into $N$ subvolumes, we
divide the sample into $N$ jackknife subsamples of equal area with the same
radial selection. Also, the mocks used in \cite{Norberg09} have smaller
volume than the ones used in this paper, which might lead to larger
uncertainties in their error estimates. It is possible that the differences
in applying the jackknife method and in the uncertainties can explain the
apparent difference between our results and those of \cite{Norberg09}.

The right panel of Figure~\ref{fig:jackknife} shows the jackknife error
estimates for the entire CMASS sample. It is encouraging that the overall
shape and magnitude of the fractional errors from the real data are in good
general agreement with those from the mocks, even though the mocks do not
intend to match the small-scale clustering. Since we do not see significant
variations in the error estimates with 100 or more jackknife samples, we
choose $N=100$ jackknife samples for the error estimation in this paper. With
an effective area of about $3,300\deg^2$ for the CMASS DR9 sample, each
excluded jackknife region then has an area of ${\sim}33\deg^2$, corresponding
to about $2.1\times 10^4$ $(\mpchi)^2$, large enough for measuring the 2PCF
in the range presented in this paper ($<50\mpchi$). The normalized jackknife
covariance matrix with $N=100$ samples, for $w_p$ of the whole CMASS sample,
is shown in Figure~\ref{fig:cov}. While the correlation coefficients for the
off-diagonal elements on small scales ($<2\mpchi$) are mostly below $0.3$,
the errors of $w_p$ on large scales are highly correlated. Therefore, when
fitting $w_p$ for scales larger than $2\mpchi$, the full covariance matrix,
not just the diagonal elements, should be taken into account.


\begin{thebibliography}{106}

\bibitem[Ahn et al.(2012)]{Ahn12} Ahn, S.-I. C. C.~P., Alexandroff, R.,
    Allende~Prieto, C., \& et~al. 2012, \apjs, 203, 21

\bibitem[Anderson et al.(2012)]{Anderson12} Anderson, L., Aubourg, E.,
    Bailey, S., et al.\ 2012, \mnras, 427, 3435

\bibitem[Baldry et al.(2004)]{Baldry04} Baldry, I.~K., Glazebrook, K.,
    Brinkmann, J., et al. 2004, \apj, 600, 681

\bibitem[Bell et al.(2004)]{Bell04} Bell, E.~F., Wolf, C., Meisenheimer, K.,
    et al. 2004, \apj, 608, 752

\bibitem[Benoist et al.(1996)]{Benoist96} Benoist, C., Maurogordato, S.,
    da~Costa, L.~N., Cappi, A., \& Schaeffer, R.
  1996, \apj, 472, 452

\bibitem[Berlind \& Weinberg(2002)]{Berlind02} Berlind, A.~A., \& Weinberg,
    D.~H. 2002, \apj, 575, 587

\bibitem[Berlind et al.(2003)]{Berlind03} Berlind, A.\ A.,  Weinberg, D.\ H.,
    Benson, A.\ J., Baugh, C.\ M., Cole, S., et al.\ 2003, \apj, 593, 1

\bibitem[Blanton et al.(2005)]{Blanton05} Blanton, M.~R., Eisenstein, D.,
    Hogg, D.~W., Schlegel, D.~J., \& Brinkmann, J.
  2005, \apj, 629, 143

\bibitem[Bolton et al.(2012)]{Bolton12} Bolton, A.~S., Schlegel, D.~J.,
    Aubourg, E., et al. 2012, \aj, 144, 144

\bibitem[Brown et al.(2008)]{Brown08} Brown, M.~J.~I., Zheng, Z., White, M.,
    et al. 2008, \apj, 682, 937

\bibitem[Budav\'ari et al.(2003)]{Budavari03} Budav\'ari, T., Connolly,
    A.~J., Szalay, A.~S., et al. 2003, \apj, 595, 59

\bibitem[Cannon et al.(2006)]{Cannon06} Cannon, R., Drinkwater, M., Edge, A.,
    et al. 2006, \mnras, 372, 425

\bibitem[Chen et al.(2012)]{Chen12} Chen, Y.-M., Kauffmann, G., Tremonti,
    C.~A., et al. 2012, \mnras, 421, 314

\bibitem[Christodoulou et al.(2012)]{Christodoulou12} Christodoulou, L.,
    Eminian, C., Loveday, J., et al. 2012, \mnras, 425, 1527

\bibitem[Coil et al.(2006)]{Coil06} Coil, A.~L., Newman, J.~A., Cooper,
    M.~C., et al. 2006, \apj, 644, 671

\bibitem[Coil et al.(2008)]{Coil08} Coil, A.~L., Newman, J.~A., Croton, D.,
    et al. 2008, \apj, 672, 153

\bibitem[Conroy \& Gunn(2010)]{Conroy10} Conroy, C., \& Gunn, J.~E. 2010,
    \apj, 712, 833

\bibitem[Conroy et al.(2009)]{Conroy09} Conroy, C., Gunn, J.~E., \& White, M.
    2009, \apj, 699, 486

\bibitem[Conroy et al.(2006)]{Conroy06} Conroy, C., Wechsler, R.~H., \&
    Kravtsov, A.~V. 2006, \apj, 647, 201

\bibitem[Cooray \& Sheth(2002)]{Cooray02} Cooray, A., \& Sheth, R. 2002,
    \physrep, 372, 1

\bibitem[Davis \& Geller(1976)]{Davis76} Davis, M., \& Geller, M.~J. 1976,
    \apj, 208, 13

\bibitem[Davis et al.(1988)]{Davis88} Davis, M., Meiksin, A., Strauss, M.~A.,
    da~Costa, L.~N., \& Yahil, A. 1988,
  \apjl, 333, L9

\bibitem[Davis \& Peebles(1983)]{Davis83} Davis, M., \& Peebles, P.~J.~E.
    1983, \apj, 267, 465

\bibitem[Dawson et al.(2013)]{Dawson13} Dawson, K.~S., Schlegel, D.~J., Ahn,
    C.~P., et al.\ 2013, \aj, 145, 10

\bibitem[Eisenstein \& Hu(1998)]{Eisenstein98} Eisenstein, D.~J., \& Hu, W.
    1998, \apj, 496, 605

\bibitem[Eisenstein et al.(2001)]{Eisenstein01} Eisenstein, D.~J., Annis, J.,
    Gunn, J.~E., et al. 2001, \aj, 122, 2267

\bibitem[Eisenstein et al.(2011)]{Eisenstein11} Eisenstein, D.~J., Weinberg,
    D.~H., Agol, E., et al. 2011, \aj, 142, 72

\bibitem[Faber et al.(2007)]{Faber07} Faber, S.~M., Willmer, C.~N.~A., Wolf,
    C., et al. 2007, \apj, 665, 265

\bibitem[Feldman et al.(1994)]{Feldman94} Feldman, H.~A., Kaiser, N., \&
    Peacock, J.~A. 1994, \apj, 426, 23

\bibitem[Fry(1996)]{Fry96} Fry, J.~N. 1996, \apjl, 461, L65

\bibitem[Fukugita et al.(1996)]{Fukugita96} Fukugita, M., Ichikawa, T., Gunn,
    J.~E., et al. 1996, \aj, 111, 1748

\bibitem[Gallazzi et al.(2006)]{Gallazzi06} Gallazzi, A., Charlot, S.,
    Brinchmann, J., \& White, S.~D.~M. 2006, \mnras,
  370, 1106

\bibitem[Gunn et al.(1998)]{Gunn98} Gunn, J.~E., Carr, M., Rockosi, C., et
    al. 1998, \aj, 116, 3040

\bibitem[Gunn et al.(2006)]{Gunn06} Gunn, J.~E., Siegmund, W.~A., Mannery,
    E.~J., et al. 2006, \aj, 131, 2332

\bibitem[Guo et al.(2012)]{Guo12} Guo, H., Zehavi, I., \& Zheng, Z. 2012,
    \apj, 756, 127

\bibitem[Guo et al.(2010)]{Guo10} Guo, Q., White, S., Li, C., \&
    Boylan-Kolchin, M. 2010, \mnras, 404, 1111

\bibitem[Guzzo et al.(1997)]{Guzzo97} Guzzo, L., Strauss, M.~A., Fisher,
    K.~B., Giovanelli, R., \& Haynes, M.~P.
  1997, \apj, 489, 37

\bibitem[Hamilton(1988)]{Hamilton88} Hamilton, A.~J.~S. 1988, \apjl, 331, L59

\bibitem[Hamilton(1993)]{Hamilton93} Hamilton, A.~J.~S.\ 1993, \apj,
    417, 19

\bibitem[Jackson(1972)]{Jackson72} Jackson, J.~C.\ 1972, \mnras, 156, 1P

\bibitem[Kaiser(1984)]{Kaiser84} Kaiser, N. 1984, \apjl, 284, L9

\bibitem[Kaiser(1987)]{Kaiser87} Kaiser, N.\ 1987, \mnras, 227, 1

\bibitem[Kodama \& Arimoto(1997)]{Kodama97} Kodama, T., \& Arimoto, N. 1997,
    \aap, 320, 41

\bibitem[Komatsu et al.(2011)]{Komatsu11} Komatsu, E., Smith, K.~M., Dunkley,
    J., et al. 2011, \apjs, 192, 18

\bibitem[Kravtsov et al.(2004)]{Kravtsov04} Kravtsov, A.~V., Berlind, A.~A.,
    Wechsler, R.~H., et al. 2004, \apj, 609, 35

\bibitem[Landy \& Szalay(1993)]{Landy93} Landy, S.~D., \& Szalay, A.~S. 1993,
    \apj, 412, 64

\bibitem[Li et al.(2006)]{Li06} Li, C., Kauffmann, G., Jing, Y.~P., et al.
    2006, \mnras, 368, 21

\bibitem[Loh et al.(2010)]{Loh10} Loh, Y.-S., Rich, R.~M., Heinis, S., et al.
    2010, \mnras, 407, 55

\bibitem[Loveday et al.(1995)]{Loveday95} Loveday, J., Maddox, S.~J.,
    Efstathiou, G., \& Peterson, B.~A. 1995, \apj, 442,
  457

\bibitem[Madgwick et al.(2003)]{Madgwick03} Madgwick, D.~S., Hawkins, E.,
    Lahav, O., et al. 2003, \mnras, 344, 847

\bibitem[Manera et al.(2013)]{Manera13} Manera, M., Scoccimarro,
    R., Percival, W.~J., et al.\ 2013, \mnras, 428, 1036

\bibitem[Maraston et al.(2012)]{Maraston12} Maraston, C., Pforr, J.,
    Henriques, B.~M., et al. 2012, \mnras, in press, arXiv:1207.6114

\bibitem[Martin et al.(2007)]{Martin07} Martin, D.~C., Wyder, T.~K.,
    Schiminovich, D., et al. 2007, \apjs, 173, 342

\bibitem[Masters et al.(2011)]{Masters11} Masters, K.~L., Maraston, C.,
    Nichol, R.~C., et al. 2011, \mnras, 418, 1055

\bibitem[Meneux et al.(2006)]{Meneux06} Meneux, B., Le~F\`evre, O., Guzzo,
    L., et al. 2006, \aap, 452, 387

\bibitem[Meneux et al.(2008)]{Meneux08} Meneux, B., Guzzo, L., Garilli, B.,
    et al. 2008, \aap, 478, 299

\bibitem[Meneux et al.(2009)]{Meneux09} Meneux, B., Guzzo, L., de~la Torre,
    S., et al. 2009, \aap, 505, 463

\bibitem[Mo \& White(1996)]{Mo96} Mo, H.~J., \& White, S.~D.~M. 1996, \mnras,
    282, 347

\bibitem[Mostek et al.(2012)]{Mostek12} Mostek, N., Coil, A.~L., Cooper,
    M.~C., et al. 2012, \apj, submitted, arXiv:1210.6694

\bibitem[Norberg et al.(2009)]{Norberg09} Norberg, P., Baugh, C.~M.,
    Gazta\~naga, E., \& Croton, D.~J. 2009, \mnras, 396,
  19

\bibitem[Norberg et al.(2001)]{Norberg01} Norberg, P., Baugh, C.~M., Hawkins,
    E., et al. 2001, \mnras, 328, 64

\bibitem[Norberg et al.(2002)]{Norberg02} Norberg, P., Baugh, C.~M., Hawkins,
    E., et al.\ 2002, \mnras, 332, 827

\bibitem[Nuza et al.(2012)]{Nuza12} Nuza, S.~E., Sanchez, A.~G., Prada, F.,
    et al. 2012, \mnras, submitted, arXiv:1202.6057

\bibitem[Padmanabhan et al.(2013)]{Padmanabhan13} Padmanabhan, N., et al.
    2013, in preparation

\bibitem[Parejko et al.(2013)]{Parejko13} Parejko, J.~K., Sunayama,
    T., Padmanabhan, N., et al.\ 2013, \mnras, 429, 98

\bibitem[Peacock \& Smith(2000)]{Peacock00} Peacock, J.~A., \& Smith, R.~E.
    2000, \mnras, 318, 1144

\bibitem[Reid et al.(2012)]{Reid12} Reid, B.~A., Samushia, L., White, M., et
    al. 2012, \mnras, 426, 2719

\bibitem[Ross \& Brunner(2009)]{Ross09} Ross, A.~J., \& Brunner,
    R.~J.\ 2009, \mnras, 399, 878

\bibitem[Ross et al.(2011b)]{Ross11a} Ross, A.~J., Ho, S., Cuesta, A.~J., et
    al. 2011a, \mnras, 417, 1350

\bibitem[Ross et al.(2010)]{Ross10} Ross, A.~J., Percival,
    W.~J., \& Brunner, R.~J.\ 2010, \mnras, 407, 420

\bibitem[Ross et al.(2013)]{Ross13} Ross, A.~J., Percival,
    W.~J., Carnero, A., et al.\ 2013, \mnras, 428, 1116

\bibitem[Ross et al.(2012)]{Ross12} Ross, A.~J., Percival, W.~J.,
    S\'anchez, A.~G., et al. 2012, \mnras, 424, 564

\bibitem[Ross et al.(2011a)]{Ross11b} Ross, A.~J., Tojeiro, R., \&
    Percival, W.~J.\ 2011b, \mnras, 413, 2078

\bibitem[Ross et al.(2007)]{Ross07} Ross, N.~P., da~\^Angela, J., Shanks, T.,
    et al. 2007, \mnras, 381, 573

\bibitem[Samushia et al.(2013)]{Samushia13} Samushia, L., Reid,
    B.~A., White, M., et al.\ 2013, \mnras, 429, 1514

\bibitem[S\'anchez et al.(2012)]{Sanchez12} S\'anchez, A.~G., Sc\'occola,
    C.~G., Ross, A.~J., et al. 2012, \mnras, 425, 415

\bibitem[Sawangwit et al.(2011)]{Sawangwit11} Sawangwit, U., Shanks,
    T., Abdalla, F.~B., et al.\ 2011, \mnras, 416, 3033

\bibitem[Schlegel et al.(1998)]{Schlegel98} Schlegel, D.~J., Finkbeiner,
    D.~P., \& Davis, M. 1998, \apj, 500, 525

\bibitem[Scoccimarro et al.(2001)]{Scoccimarro01} Scoccimarro, R., Sheth,
    R.~K., Hui, L., \& Jain, B. 2001, \apj, 546, 20

\bibitem[Sc\'occola et al.(2012)]{Scoccola12} Sc\'occola, C.~G., S\'anchez,
    A.~G., Rubi\~no Martin, J.~A., et al. 2012, \mnras, submitted,
arXiv:1209.1394

\bibitem[Scoville et al.(2007)]{Scoville07} Scoville, N., Abraham, R.~G.,
    Aussel, H., et al. 2007, \apjs, 172, 38

\bibitem[Seljak(2000)]{Seljak00} Seljak, U. 2000, \mnras, 318, 203

\bibitem[Shen et al.(2012)]{Shen12} Shen, Y., McBride, C.~K.,
    White, M., et al.\ 2012, \apj, submitted, arXiv:1212.4526

\bibitem[Skelton et al.(2009)]{Skelton09} Skelton, R.~E., Bell, E.~F., \&
    Somerville, R.~S. 2009, \apjl, 699, L9

\bibitem[Skibba \& Sheth(2009)]{Skibba09a} Skibba, R.~A., \& Sheth, R.~K.
    2009, \mnras, 392, 1080

\bibitem[Skibba et al.(2009)]{Skibba09b} Skibba, R.~A., Bamford, S.~P.,
    Nichol, R.~C., et al. 2009, \mnras, 399, 966

\bibitem[Smee et al.(2012)]{Smee12} Smee, S., Gunn, J.~E., Uomoto, A., et al.
    2012, \aj, submitted, arXiv:1208.2233

\bibitem[Smith et al.(2003)]{Smith03} Smith, R.~E., Peacock, J.~A., Jenkins,
    A., et al. 2003, \mnras, 341, 1311

\bibitem[Springel et al.(2005)]{Springel05} Springel, V., White, S.~D.~M.,
    Jenkins, A., et al. 2005, \nat, 435, 629

\bibitem[Strateva et al.(2001)]{Strateva01} Strateva, I., Ivezi\'c, v.~Z.,
    Knapp, G.~R., et al. 2001, \aj, 122, 1861

\bibitem[Swanson et al.(2013)]{Swanson13} Swanson, M.~E.~C., et al. 2013, in
    preparation

\bibitem[Swanson et al.(2008)]{Swanson08} Swanson, M.~E.~C., Tegmark, M.,
    Blanton, M., \& Zehavi, I. 2008, \mnras, 385,
  1635

\bibitem[Tinker \& Wetzel(2010)]{Tinker10} Tinker, J.~L., \& Wetzel, A.~R.
    2010, \apj, 719, 88

\bibitem[Tojeiro et al.(2011)]{Tojeiro11} Tojeiro, R., Percival, W.~J.,
    Heavens, A.~F., \& Jimenez, R. 2011, \mnras, 413,
  434

\bibitem[Tojeiro et al.(2012a)]{Tojeiro12a} Tojeiro, R., Percival, W.~J.,
    Brinkmann, J., et al. 2012a, \mnras, 424, 2339

\bibitem[Tojeiro et al.(2012b)]{Tojeiro12b} Tojeiro, R., Percival, W.~J.,
    Wake, D.~A., et al. 2012b, \mnras, 424, 136

\bibitem[Wang et al.(2007)]{Wang07} Wang, Y., Yang, X., Mo, H.~J., \& van den
    Bosch, F.~C.\ 2007, \apj, 664, 608

\bibitem[Wake et al.(2008)]{Wake08} Wake, D.~A., Sheth, R.~K.,
    Nichol, R.~C., et al.\ 2008, \mnras, 387, 1045

\bibitem[Wake et al.(2011)]{Wake11} Wake, D.~A., Whitaker, K.~E., Labb{\'e},
    I., et al.\ 2011, \apj, 728, 46

\bibitem[Weinberg et al.(2012)]{Weinberg12} Weinberg, D.~H., Mortonson,
    M.~J., Eisenstein, D.~J., et al. 2012, arXiv:1201.2434

\bibitem[Whitaker et al.(2010)]{Whitaker10} Whitaker, K.~E., van Dokkum,
    P.~G., Brammer, G., et al. 2010, \apj, 719, 1715

\bibitem[White et al.(2007)]{White07} White, M., Zheng, Z., Brown, M.~J.~I.,
    Dey, A., \& Jannuzi, B.~T. 2007, \apjl,
  655, L69

\bibitem[White et al.(2011)]{White11} White, M., Blanton, M., Bolton, A., et
    al. 2011, \apj, 728, 126

\bibitem[Yang et al.(2005)]{Yang05} Yang, X., Mo, H.~J., Jing, Y.~P., \&
    van~den Bosch, F.~C. 2005, \mnras, 358,
  217

\bibitem[Yang et al.(2003)]{Yang03} Yang, X., Mo, H.~J., \& van~den Bosch,
    F.~C. 2003, \mnras, 339, 1057

\bibitem[York et al.(2000)]{York00} York, D.~G., Adelman, J., Anderson, Jr.,
    J.~E., et al. 2000, \aj, 120, 1579

\bibitem[Zehavi et al.(2002)]{Zehavi02} Zehavi, I., Blanton, M.~R., Frieman,
    J.~A., et al. 2002, \apj, 571, 172

\bibitem[Zehavi et al.(2004)]{Zehavi04} Zehavi, I., Weinberg, D.~H., Zheng,
    Z., et al. 2004, \apj, 608, 16

\bibitem[Zehavi et al.(2005a)]{Zehavi05a} Zehavi, I., Eisenstein, D.~J.,
    Nichol, R.~C., et al. 2005a, \apj, 621, 22

\bibitem[Zehavi et al.(2005b)]{Zehavi05b} Zehavi, I., Zheng, Z., Weinberg,
    D.~H., et al. 2005b, \apj, 630, 1

\bibitem[Zehavi et al.(2011)]{Zehavi11} Zehavi, I., Zheng, Z.,
    Weinberg, D.~H., et al.\ 2011, \apj, 736, 59  [Z11]

\bibitem[Zheng et al.(2007)]{Zheng07} Zheng, Z., Coil, A.~L., \& Zehavi, I.
    2007, \apj, 667, 760

\bibitem[Zheng et al.(2009)]{Zheng09} Zheng, Z., Zehavi, I., Eisenstein,
    D.~J., Weinberg, D.~H., \& Jing, Y.~P. 2009,
  \apj, 707, 554

\bibitem[Zheng et al.(2005)]{Zheng05} Zheng, Z., Berlind, A.~A., Weinberg,
    D.~H., et al. 2005, \apj, 633, 791

\end{thebibliography}

\end{document}